\documentclass[twocolumn, tighten, times, astrosymb]{aastex631}

\usepackage{xcolor}

\newcommand\dracoii{Draco~{\sc II}}
\newcommand\jwst{\textit{JWST}}
\newcommand\hst{\textit{HST}}
\newcommand\msun{$M_{\odot}$}
\newcommand\mstar{$M_{\star}$}

\accepted{February 2, 2024}

\submitjournal{ApJS}

\shorttitle{JWST Resolved Stellar Populations V}
\shortauthors{Weisz et al.}

\begin{document}

\title{The JWST Resolved Stellar Populations Early Release Science Program V. \\ DOLPHOT Stellar Photometry for NIRCam and NIRISS}

\correspondingauthor{Daniel R. Weisz}
\email{dan.weisz@berkeley.edu}

\author[0000-0002-6442-6030]{Daniel R. Weisz}
\affiliation{Department of Astronomy, University of California, Berkeley, CA 94720, USA}

\author[0000-0001-8416-4093]{Andrew E. Dolphin}
\affiliation{Raytheon, 1151 E. Hermans Rd.,
Tucson, AZ 85756}
\affiliation{Steward Observatory, University of Arizona, 933 N. Cherry Avenue, Tucson, AZ 85719, USA}

\author[0000-0002-1445-4877]{Alessandro Savino}
\affiliation{Department of Astronomy, University of California, Berkeley, CA 94720, USA}

\author[0000-0001-5538-2614]{Kristen B. W. McQuinn}
\affiliation{Department of Physics and Astronomy, Rutgers, the State University of New Jersey,  136 Frelinghuysen Road, Piscataway, NJ 08854, USA}

\author[0000-0002-8092-2077]{Max J. B. Newman}
\affiliation{Department of Physics and Astronomy, Rutgers, the State University of New Jersey,  136 Frelinghuysen Road, Piscataway, NJ 08854, USA}

\author[0000-0002-7502-0597]{Benjamin F. Williams}
\affiliation{Department of Astronomy, University of Washington, Box 351580, U.W., Seattle, WA 98195-1580, USA}

\author[0000-0002-3204-1742]{Nitya Kallivayalil}
\affiliation{Department of Astronomy, University of Virginia, 530 McCormick Road, Charlottesville, VA 22904, USA}

\author[0000-0003-2861-3995]{Jay Anderson}
\affiliation{Space Telescope Science Institute, 3700 San Martin Drive, Baltimore, MD 21218, USA}

\author[0000-0003-4850-9589]{Martha L. Boyer}
\affiliation{Space Telescope Science Institute, 3700 San Martin Drive, Baltimore, MD 21218, USA}

\author[0000-0001-6464-3257]{Matteo Correnti}
\affiliation{INAF Osservatorio Astronomico di Roma, Via Frascati 33, 00078, Monteporzio Catone, Rome, Italy}
\affiliation{ASI-Space Science Data Center, Via del Politecnico, I-00133, Rome, Italy}

\author[0000-0002-7007-9725]{Marla C. Geha}
\affiliation{Department of Astronomy, Yale University, New Haven, CT 06520, USA}

\author[0000-0002-4378-8534]{Karin M. Sandstrom}
\affiliation{Department of Astronomy \& Astrophysics, University of California San Diego, 9500 Gilman Drive, La Jolla, CA 92093, USA}

\author[0000-0003-0303-3855]{Andrew A. Cole}
\affiliation{School of Natural Sciences, University of Tasmania, Private Bag 37, Hobart, Tasmania 7001, Australia}

\author[0000-0003-1634-4644]{Jack T. Warfield}
\affiliation{Department of Astronomy, University of Virginia, 530 McCormick Road, Charlottesville, VA 22904, USA}

\author[0000-0003-0605-8732]{Evan D. Skillman}
\affiliation{University of Minnesota, Minnesota Institute for Astrophysics, School of Physics and Astronomy, 116 Church Street, S.E., Minneapolis,
MN 55455, USA}

\author[0000-0002-2970-7435]{Roger E. Cohen}
\affiliation{Department of Physics and Astronomy, Rutgers, the State University of New Jersey,  136 Frelinghuysen Road, Piscataway, NJ 08854, USA}

\author[0000-0002-1691-8217]{Rachael Beaton}
\affiliation{Space Telescope Science Institute, 3700 San Martin Drive, Baltimore, MD 21218, USA}
\affiliation{Department of Astrophysical Sciences, Princeton University, 4 Ivy Lane, Princeton, NJ 08544, USA}
\affiliation{The Observatories of the Carnegie Institution for Science, 813 Santa Barbara St., Pasadena, CA 91101, USA}

\author{Alessandro Bressan}
\affiliation{SISSA, Via Bonomea 265, 34136 Trieste, Italy}

\author[0000-0002-5480-5686]{Alberto Bolatto}
\affiliation{Department of Astronomy, University of Maryland, College Park, MD 20742, USA}
\affiliation{Joint Space-Science Institute, University of Maryland, College Park, MD 20742, USA}

\author[0000-0002-9604-343X]{Michael Boylan-Kolchin}
\affiliation{Department of Astronomy, The University of Texas at Austin, 2515 Speedway, Stop C1400, Austin, TX 78712-1205, USA}

\author[0000-0002-0372-3736]{Alyson M. Brooks}
\affiliation{Department of Physics and Astronomy, Rutgers, the State University of New Jersey,  136 Frelinghuysen Road, Piscataway, NJ 08854, USA}
\affiliation{Center for Computational Astrophysics, Flatiron Institute, 162 Fifth Avenue, New York, NY 10010, USA}

\author[0000-0003-4298-5082]{James S. Bullock}
\affiliation{Department of Physics and Astronomy, University of California, Irvine, CA 92697 USA}

\author[0000-0002-1590-8551]{Charlie Conroy}
\affiliation{Center for Astrophysics | Harvard \& Smithsonian, Cambridge, MA, 02138, USA}

\author[0000-0003-1371-6019]{Michael C. Cooper}
\affiliation{Department of Physics and Astronomy, University of California, Irvine, CA 92697 USA}

\author[0000-0002-1264-2006]{Julianne J. Dalcanton}
\affiliation{Department of Astronomy, University of Washington, Box 351580, U.W., Seattle, WA 98195-1580, USA}
\affiliation{Center for Computational Astrophysics, Flatiron Institute, 162 Fifth Avenue, New York, NY 10010, USA}

\author[0000-0002-4442-5700]{Aaron L. Dotter}
\affiliation{Department of Physics and Astronomy, Dartmouth College, 6127 Wilder Laboratory, Hanover, NH 03755, USA}

\author[0000-0002-3122-300X]{Tobias K. Fritz}
\affiliation{Department of Astronomy, University of Virginia, Charlottesville, 530 McCormick Road, VA 22904-4325, USA}

\author[0000-0001-9061-1697]{Christopher T. Garling}
\affiliation{Department of Astronomy, University of Virginia, 530 McCormick Road, Charlottesville, VA 22904, USA}

\author[0000-0002-5581-2896]{Mario Gennaro}
\affiliation{Space Telescope Science Institute, 3700 San Martin Drive, Baltimore, MD 21218, USA}
\affiliation{The William H. Miller {\sc III} Department of Physics \& Astronomy, Bloomberg Center for Physics and Astronomy, Johns Hopkins University, 3400 N. Charles Street, Baltimore, MD 21218, USA}

\author[0000-0003-0394-8377]{Karoline M. Gilbert}
\affiliation{Space Telescope Science Institute, 3700 San Martin Drive, Baltimore, MD 21218, USA}
\affiliation{The William H. Miller {\sc III} Department of Physics \& Astronomy, Bloomberg Center for Physics and Astronomy, Johns Hopkins University, 3400 N. Charles Street, Baltimore, MD 21218, USA}

\author[0000-0002-6301-3269]{Leo Girardi}
\affiliation{Padova Astronomical Observatory, Vicolo dell'Osservatorio 5, Padova, Italy}

\author[0000-0002-9280-7594]{Benjamin D. Johnson}
\affiliation{Center for Astrophysics | Harvard \& Smithsonian, Cambridge, MA, 02138, USA}

\author[0000-0001-6421-0953]{Cliff Johnson}
\affiliation{Center for Interdisciplinary Exploration and Research in Astrophysics (CIERA) and Department of Physics and Astronomy, Northwestern University, 1800 Sherman Avenue, Evanston, IL 60201, USA}

\author[0000-0001-9690-4159]{Jason Kalirai}
\affiliation{John Hopkins Applied Physics Laboratory, 11100 Johns Hopkins Road, Laurel, MD 20723, USA}

\author[0000-0001-6196-5162]{Evan N. Kirby}
\affiliation{Department of Physics, University of Notre Dame, Notre Dame, IN 46556, USA}

\author[0000-0002-1172-0754]{Dustin Lang}
\affiliation{Perimeter Institute for Theoretical Physics, Waterloo, ON N2L 2Y5, Canada}

\author[0000-0002-9137-0773]{Paola Marigo}
\affiliation{Department of Physics and Astronomy G. Galilei, University of Padova, Vicolo dell’Osservatorio 3, I-35122, Padova, Italy}

\author[0000-0002-3188-2718]{Hannah Richstein}
\affiliation{Department of Astronomy, University of Virginia, 530 McCormick Road, Charlottesville, VA 22904, USA}

\author[0000-0002-3569-7421]{Edward F. Schlafly}
\affiliation{Space Telescope Science Institute, 3700 San Martin Dr., Baltimore, MD 21218, USA}

\author[0000-0002-9599-310X]{Erik J. Tollerud}
\affiliation{Space Telescope Science Institute, 3700 San Martin Drive, Baltimore, MD 21218, USA}

\author[0000-0003-0603-8942]{Andrew Wetzel}
\affiliation{Department of Physics and Astronomy, University of California, Davis, CA 95616, USA}

\begin{abstract}

We present NIRCam and NIRISS modules for DOLPHOT, a widely-used crowded field stellar photometry package. We describe details of the modules including pixel masking, astrometric alignment, star finding, photometry, catalog creation, and artificial star tests (ASTs). We tested these modules using NIRCam and NIRISS images of M92 (a Milky Way globular cluster), Draco II (an ultra-faint dwarf galaxy), and WLM (a star-forming dwarf galaxy). DOLPHOT's photometry is highly precise and the color-magnitude diagrams are deeper and have better definition than anticipated during original program design in 2017. The primary systematic uncertainties in DOLPHOT's photometry arise from mismatches in the model and observed point spread functions (PSFs) and aperture corrections, each contributing $\lesssim0.01$~mag to the photometric error budget. Version 1.2 of WebbPSF models, which include charge diffusion and interpixel capacitance effects, significantly reduced PSF-related uncertainties. We also observed minor ($\lesssim0.05$~mag) chip-to-chip variations in NIRCam's zero points, which will be addressed by the JWST flux calibration program. Globular cluster observations are crucial for photometric calibration. Temporal variations in the photometry are generally $\lesssim0.01$~mag, although rare large misalignment events can introduce errors up to 0.08 mag. We provide recommended DOLPHOT parameters, guidelines for photometric reduction, and advice for improved observing strategies. Our ERS DOLPHOT data products are available on MAST, complemented by comprehensive online documentation and tutorials for using DOLPHOT with JWST imaging data.

\end{abstract}


\section{Introduction} \label{sec:intro}

\jwst\ has the potential to resolve millions of stars in thousands of galaxies out to large distances (e.g., $D\sim100$~Mpc).  Such data will enable new foundational science in a broad range of areas such as the cosmic distance ladder and local $H_0$ measurements, reionization,  globular cluster formation, dark matter, the stellar initial mass function, galaxy assembly, the effects of rare red stars that can effect the spectral energy distributions (SEDs) of galaxies at all cosmic epochs, and much more (e.g., see discussion in \citealt{weisz2023a}).

Much of this science comes from observations of resolved stars in crowded fields.  In crowded fields, neighboring stars have overlapping point spread functions (PSFs), which can lead to confusion over the number of stars and their relative contributions to the observed flux in a given pixel.  Recovering accurate and precise photometry for large numbers of stars in the limit of modest-to-severe crowding is technically daunting and requires highly optimized observations and sophisticated analysis tools \citep[e.g.,][]{dalcanton2012b, williams2014}.  

Fortunately, crowded field stellar photometry is a mature field based on a rich history of development dating back nearly $\sim50$ years.  Early crowded field photometry routines combined pioneering work on photoelectric detectors with innovative approaches to simultaneously modeling the stellar light profiles of adjacent stars, resulting in a number of codes in the 1980s that could photometer thousands of stars in a field \citep[e.g.,][]{buonanno1979, tody1980, stryker1983, lupton1986, penny1986, schechter1993}. A major achievement of this era was the creation of the legacy software package DAOPHOT \citep{stetson1987}.\footnote{As discussed by \citet{stetson1987}, DAOPHOT was a version of the photometric routine POORMAN written by Mould \& Shortridge that was improved to handle a higher density of stars.  Unfortunately, there is no bibliographic record for POORMAN.}

Continued improvements in crowded field photometry were catalyzed by the launch of the \textit{Hubble Space Telescope} (\hst).  Via the Hubble Key Project aimed at measuring $H_0$,  several independent photometric routines were developed to gauge systematics in the photometry (e.g., \citealt{stetson1994}; and see discussion in \citealt{freedman2001}).  Similarly, the ground-breaking sensitivity and precision of \hst/WFPC2, along with its notoriously undersampled PSF, motivated the development of specialized photometric and astrometric routines aimed at dense stellar fields \citep[e.g.,][]{holtzman1995, lauer1999, anderson2000, dolphin2000b}. More recently, stellar surveys of crowded fields such as the Galactic plane and M31 have provided important gains in the speed and flexibility of crowded field codes \citep[e.g.,][]{dalcanton2012b, schlafly2018}.  In the context of nearby galaxies, the Panchromatic Hubble Andromeda Treasury (PHAT) survey provided substantial new additions (e.g., simultaneous multi-camera, multi-wavelength crowded field photometry) to DOLPHOT \citep{dolphin2000b, dolphin2016}, a crowded field photometric package that has produced photometry for millions of stars in hundreds of galaxies in the Local Group (LG) and Local Volume \citep[e.g.,][]{holtzman2006, rizzi2007, weisz2008, dalcanton2009, mcquinn2010, Radburn-Smith:2011qf, dalcanton2012, dalcanton2012b, williams2014, jang2017, mcquinn2017, skillman2017, sabbi2018, anand2021, jang2021, williams2021, lee2022, savino2022, riess2023, williams2023}.

A main goal of our \jwst\ Resolved Stellar Populations Early Release Science is to provide the astronomy community with an easy-to-use and efficient means for performing crowded field photometry on \jwst\ imaging that will ultimately help to realize \jwst's full potential for the resolved Universe.  Specifically, we have developed modules for DOLPHOT that are tailored to the characteristics of NIRCam and NIRISS, which are important imaging instruments for studies of resolved stellar populations with \jwst. DOLPHOT is a well-tested, widely-used, and publicly available package that already supports modules specific to several \textit{HST} cameras (WFPC2, ACS, WFC3/UVIS and IR), has been a testing ground for the \textit{Roman Space Telescope}, and includes general purpose routines that can be used on virtually any images of resolved stars.  The addition of \jwst\ modules will enable a wide array of \jwst-specific and cross-facility (e.g., \jwst\ and \hst) science, some of which has already been demonstrated using early versions of our \jwst\ DOLPHOT modules \citep[e.g.,][]{chen2023, lee2024, lee2023, mcquinn2023, riess2023, vandyk2023, warfield2023, peltonen2024, li2024}.

\begin{table*}
    
    \caption{A summary of our \jwst\ Early Release Science (ERS) observations taken in 2022. Although we acquired 4 exposures for M92, the 3rd exposure produces poor photometry due to larger-than-normal jitter in the telescope stability that occurred only during this exposure.  We examine this issue in Appendix \ref{sec:m92_3rd}. The first entry for M92 in this table reflects all observations taken, while the second entry is without the 3rd exposure.  More details on the exact observations (e.g., dither pattern, readout mode) are given in \citet{weisz2023a} and are available in our public Phase II file in APT. \label{tab:obs}}
    \centering
    \begin{tabular}{lccccccc}

    \toprule
    Target & Date & Camera & Filter & $t_{exp}$ [s] & Groups & Integrations & Dithers\\
    \toprule
    M92 & June 20--21 & NIRCam & F090W/F277W & 1245.465 & 6 & 1 & 4\\
     & & NIRCam & F150W/F444W & 1245.465 & 6 & 1 & 4\\
     & & NIRISS & F090W & 1245.465 & 7 & 1 & 4\\
     & & NIRISS & F150W & 1245.465 & 7 & 1 & 4\\
\hline
M92 (no 3rd exp) & June 20--21 & NIRCam & F090W/F277W & 934.099 & 6 & 1 & 3\\
     & & NIRCam & F150W/F444W & 934.099 & 6 & 1 & 3\\
     & & NIRISS & F090W & 934.099 & 7 & 1 & 3\\
     & & NIRISS & F150W & 934.099 & 7 & 1 & 3\\
     \hline
     Draco~{\sc II} & July 3 & NIRCam & F090W/F480M & 11810.447 & 7 & 4 & 4\\
     & & NIRCam & F150W/F360M & 5883.75 & 7 & 2 & 4\\
     & & NIRISS & F090W & 11123.294 & 9 & 7 & 4\\
     & & NIRISS & F150W & 5883.75 & 10 & 3 & 4\\
     WLM & July 23--24 & NIRCam & F090W/F430M & 30492.427 & 8 & 9 & 4\\
     & & NIRCam & F150W/F250M & 23706.788 & 8 & 7 & 4\\
     & & NIRISS & F090W & 26670.137 & 17 & 9 & 4\\
     & & NIRISS & F150W & 19841.551 & 19 & 6 & 4\\
     \hline
    \toprule

    \end{tabular}
\end{table*}

In this paper, we describe the NIRCam and NIRISS stellar photometry modules for DOLPHOT.  DOLPHOT's underlying algorithms are already well-documented in the literature, along with rigorous tests of their accuracy in a variety of crowded  and uncrowded fields \citep[e.g.,][]{Radburn-Smith:2011qf, williams2014, williams2023}.  Accordingly, our focus is on describing the details specific to the DOLPHOT NIRCam and NIRISS modules and providing examples of its application to the three Early Release Science (ERS) targets in the Local Group:  M92, \dracoii, and WLM.  As described in \citet{weisz2023a}, these targets, and the associated observing strategies, were carefully selected to benchmark the development of DOLPHOT in a variety of regimes that we anticipate will be common for resolved star science, and thus need to be vetted for study with \jwst. This paper is designed to describe the modules and provide examples of their application to the ERS data.  As part of the ERS program, we have created an extensive set of deliverables including online documentation and data products that allow interested readers to reduce ERS data identically to what is done in this paper, as well as explore the various aspects of the data for their own purposes (e.g., to  customize catalog culling criteria). Essential DOLPHOT input and output data associated with the photometric reductions in this paper are hosted as high-level science products on MAST\footnote{\url{https://archive.stsci.edu/hlsp/jwststars/}}, while step-by-step guides for our DOLPHOT reductions can be found on our DOLPHOT documentation page\footnote{\url{https://dolphot-jwst.readthedocs.io}}.

This paper is organized as follows.   We summarize the ERS observations in \S \ref{sec:obs}.  In \S \ref{sec:dolphot}, we describe the DOLPHOT NIRCam and NIRISS modules and provide a general outline of how to apply these new modules to \jwst\ imaging in order to produce stellar catalogs.  We illustrate the application of these modules to ERS data in \S \ref{sec:ers_photometry}.  In \S \ref{sec:discussion}, we examine the time variability of the PSF and compare the DOLPHOT SNR estimates with expectations from the \jwst\ ETC.  Finally, we summarize the paper and highlight future areas for improvement in \S \ref{sec:conclude}.

\section{Observations}
\label{sec:obs}

Extensive details of our ERS survey and observations are provided in \citet{weisz2023a}.  Here, we briefly summarize the observations and list their basic characteristics in Table~\ref{tab:obs}.

In June and July, 2022 our program acquired NIRCam and NIRSS imaging of three LG targets: globular cluster M92, ultra-faint dwarf galaxy \dracoii, and LG star-forming dwarf galaxy WLM.  These targets were selected to satisfy a number of science and technical goals including the development and testing of DOLPHOT in a variety of scenes (e.g., crowded and uncrowded fields, varying surface brightness, varying degrees of saturation, a representative set of wide and medium filters).  In all cases, the NIRCam fields were placed centrally on each target with locations and orientations set to maximize overlap with archival \hst\ imaging and schedulability early in the ERS window.  The NIRISS fields were acquired in parallel.  Table \ref{tab:obs} summarizes basic characteristics of our NIRCam and NIRISS observations.

In the process of analyzing our M92 data, we found that the third exposure of M92 appears to be ``corrupt'' in the sense that although the 3rd exposure of M92 visually looks fine, it results in remarkably poor photometry for both NIRCam and NIRISS, despite extensive efforts to fix it.  We therefore have excluded it from the DOLPHOT reductions in this paper.  A later analysis of the fine guidance sensor data revealed instabilities in the telescope only during this exposure.  We discuss details of the third exposure in Appendix \ref{sec:m92_3rd}.

Our observations of WLM were designed to sample RR Lyrae light curves.  However, the default \jwst\ reduction pipeline is currently not capable of producing the time series images for short period observations needed to extract flux as a function of time.   The default pipeline currently only provides time series data when the time series observation (TSO) mode is used.  We were unable to use TSO mode because it prohibits dithering. While it is possible to modify the \jwst\ pipeline to produce the time series images necessary for short period variable analysis, it is a topic beyond the scope of this paper.  Instead, a Cycle 2 archival proposal undertaken by some members of our team (AR-03248; PI Skillman) is developing the documentation and tools needed to recover variable star light curves from NIRCam and NIRISS imaging with DOLPHOT.  These will be made publicly available upon completion.

Finally, the NIRISS observations of \dracoii, were located at several half-light radii from the galaxy.  As far as we can tell, the field is consistent with being blank, i.e., no obvious galaxy member stars, and we do not analyze or discuss this field in the paper.

\section{NIRCam and NIRISS Photometry with DOLPHOT}
\label{sec:dolphot}

In this section, we provide an overview of the new DOLPHOT NIRCam and NIRISS modules with application to imaging from our ERS program.  The core workings of DOLPHOT, along with extensive tests of its functionality and reliability are well-documented in the literature \citep[e.g.,][]{dolphin2000b, dalcanton2012, dalcanton2012b, williams2014, dolphin2016, williams2021}.  Here, we will not re-visit these details. Instead, we focus on modifications made to DOLPHOT for incorporating NIRCam and NIRISS imaging into its existing framework.

\subsection{Overview}
\label{sec:dolphot_overview}

As input, DOLPHOT takes a reference image, a list of science images, and several dozen input parameters with user defined values. DOLPHOT astrometically aligns all the science images to a reference image.  It then performs simultaneously, multi-wavelength photometry on all science images by fitting PSF models to the signal-to-noise ratio (SNR) peaks (and any neighboring SNR peaks in crowded fields) it detects in each of the science images.  The result of this process is a set of photometric measurements for all detected objects in all science images.  Stellar catalogs are created by culling the main source catalog using criteria such as SNR, how compact/extended sources are, etc., which we detail in \S \ref{sec:cull}.

For \jwst, the science images we use are the stage 2 \texttt{cal} files, which are calibrated single exposure science images produced by the \jwst\ pipeline.  The  references images are \texttt{I2D} files, which are resampled, stacked science images, akin to drizzled images with \hst. It is possible to use \texttt{cal} as reference images as well.  The \jwst\ pipeline also makes available \texttt{crf} images, which are like \texttt{cal} science images, but with cosmic ray flags applied.  Generally, we have found astrometric alignment of the \texttt{crf} images in DOLPHOT to be worse than \texttt{cal} images.

The reference image is used only to align each of the science images. Our general recommendation for a reference image is to select the deepest image available.  All images used in this analysis were created by the standard STScI pipeline and downloaded from MAST\footnote{The specific observations used in his paper can be accessed via \dataset[DOI: 10.17909/71kb-ga31]{https://doi.org/10.17909/71kb-ga31}.}.

In this paper, we use FITS images with the following \jwst\ pipeline versioning information \texttt{CAL\_VER}$=$1.11.4, \texttt{CRDS\_VER}$=$11.17.2, and \texttt{CRDS\_CTX}$=$jwst\_p1147.pmap. This version includes updates to the chip-to-chip zero points, the switch from Vega to Sirus as a reference star, and updated flat fields released in late 2023.

DOLPHOT requires only a few pre-processing steps for all images. These steps include masking bad pixels (e.g., cosmic rays, hot pixels) based on the data quality (DQ) flags, multiplying by the camera specific pixel area masks, and making initial estimates of the sky and the positions of bright stars for alignment. 

Following pre-processing, DOLPHOT aligns each science image to the reference image.  The quality of the alignment is determined by several factors including the number of bright stars available, depth of the reference image, the fidelity of the provided WCS information, SNRs, relative orientations (e.g., images with large rotations may be harder to align), stars in common between the reference and science image (e.g., images taken in very different filters, such as ultra-violet and IR, may be hard to align as they may have few sources in common). 

With all images aligned, DOLPHOT searches for objects to photometer by iteratively identifying signal-to-noise ratio peaks in the stack of science images. DOLPHOT measures the fluxes of each object by simultaneously fitting a PSF model, and a local background model, to the target object plus all neighboring objects within a user specified radius.  

Upon completion of photometry, DOLPHOT provides extensive output including its position on the reference image and the flux and a number of quality assessment metrics (e.g., $\chi^2$, shape of the star's light profile relative to the PSF) for each star in each image.  It also provides combined fluxes and magnitudes for each object from which stellar catalogs are usually constructed.

Characterizing uncertainties for crowded field stellar photometry requires artificial star tests (ASTs).  ASTs are synthetic stars with known positions and magnitudes that are inserted into real \jwst\ science images and then recovered by DOLPHOT. It is well-established that the difference in input and recovered flux for ASTs provides a more realistic accounting of photometric uncertainties than the Poisson noise that is reported by the crowded field photometric process alone \citep[e.g.,][]{stetson1988}.

\subsection{Pre-processing steps}

Prior to running DOLPHOT, pre-processing is required in order to convert the data to a format suitable for PSF-fitting photometry.  For the case of NIRCAM and NIRISS data, steps in this process are as follows (using the \texttt{nircammask} and \texttt{nirissmask} utilities, respectively):

\begin{itemize}
    \item Mask out bad or saturated pixels.  At the time of this writing, bad pixels on \texttt{cal} and \texttt{crf}  images are identified by having an SCI array value of NaN; previous versions of the pipeline have used SCI array values of exactly 0.  The mask utilities will correctly interpret either approach.  Additionally, saturated pixels in \texttt{cal} and \texttt{crf}  images are identified by having a DQ array flag with a value of 2.  Bad pixels on \texttt{I2D}  images are identified by having a WHT array value of exactly 0.
    
    \item Convert from the default calibration of MJy/sr to DN (data number).  This is performed by dividing all pixel values by the FITS keyword PHOTMJSR, and subsequently multiplying by the exposure time (FITS keyword EFFEXPTM)\footnote{We note that as of this writing, the actual time the telescope spends collecting data slightly differs from the exposure time in the FITS keywords, such as EFFEXPTM.  In practice, the ramp fitting procedure begins at the end of the first group. But currently, the FITS exposure time keywords are based on when the first group starts.  This effect is generally subtle, i.e., the impact on SNR is typically $\lesssim1$\%, but it is now factored into DOLPHOT for accuracy and completeness.}.  An additional step for \texttt{cal}  and \texttt{crf}  images is to multiply pixel values by the pixel area map (AREA array).

    \item Readout noise and gain values are also saved into the FITS file.  Details of the pedigree of that data are available in the \texttt{versions.txt} files that are included with the  NIRCAM and NIRISS DOLPHOT modules.

\end{itemize}

The result of the pre-processing step is an image in units of DN, along with FITS keywords for gain and readout noise, allowing PSF-fitting photometry to run.

As of this writing, for the purpose of this ERS program, there was no need to incorporate information from the Advanced Scientific Data Format (ASDF) metadata provided for \jwst\ images \citep{greenfield2015}.  ASDF has the ability to host more detailed metadata than the standard FITS format, such as improved WCS information.  However, at this time  all necessary information (astrometry information, photometric calibrations, pixel areas, etc.) for DOLPHOT are available via FITS header keywords.  If that situation changes in the future, we will explore writing an ASDF reader for DOLPHOT.

\subsection{Alignment}

The first step in DOLPHOT’s reduction process is to align all of the original-sampling (\texttt{cal} or \texttt{crf}) images to a common reference frame.  Normally, the common reference is an \texttt{I2D} file.  In principle, if no dithering was used in the observations, one of the \texttt{cal} or \texttt{crf} images could be used as a reference image.   In practice, typically the deepest \texttt{I2D} file makes for the best reference image as it allows for the most star matches with the science images leading to better astrometric alignment.  We note that DOLPHOT does not re-sample or re-bin any images, which can lead to issues in conservation of intensity, for example.  Instead, DOLPHOT only performs photometry on the original, non-drizzled, science images provided by the \jwst\ pipeline.

A key difference between the \jwst\ modules and the previously released DOLPHOT \hst\ modules is that DOLPHOT does not apply distortion corrections based on DOLPHOT's internal model.  With the \hst\ modules, common practice is to use the astrometric data in the header but not the distortion.  Using \hst\ with DOLPHOT's internal distortion model requires the parameter setting UseWCS=1, for which DOLPHOT estimates only the shift, scale, and rotation of each science image relative to the reference.  For \jwst\, the astrometric data included in the FITS header is used for both alignment and application of distortion corrections, requiring a setting of UseWCS=2.  In this case, DOLPHOT estimates a full distortion solution, which is beyond the shift, scale, and rotation typically used by DOLPHOT on \hst\ images.  As of this writing, NIRCAM data are provided with 3rd order SIP polynomials, while NIRISS data are provided with 4th order polynomials. DOLPHOT currently processes up to 5th order polynomials, so it can handle additional fidelity in the astrometry data, should it become available in the pipeline or added by offline astrometry tools (e.g., astrometry.net; \citealt{lang2010}).

\subsection{Star Detection and Photometry} 
\label{sec:dolphot_photometry}
 
As with previously available DOLPHOT modules, photometry proceeds once the images are aligned.  An initial pass (and usually multiple passes) detects peaks in the SNR map across the reference image. Photometry is performed at each peak to attempt to identify a point source.  As with all previous versions of DOLPHOT, the user can adjust parameters to alter the noise models, photometry modes (e.g., aperture vs. PSF-fitting), sky fitting method, etc.  Our internal testing on the ERS data resulted in a set of recommended DOLPHOT parameters, which are listed in Table \ref{tab:setup}.  We discuss the process by which we determined these parameters in more detail in \S \ref{sec:optimizing}.

A standard feature of DOLPHOT is to make adjustments to the model PSFs to improve the fit quality and photometry.  As discussed in \citet{dolphin2000b}, mismatches between the shape of the model and the true PSF contribute to the photometric error budget, which can grow large for faint sources.  As part of its normal operation, DOLPHOT measures a PSF residual image relative to the pre-calculated PSF model library.  It then makes adjustments to the model PSFs based on comparisons to bright stars in the field to improve agreement between the model and observed PSFs.  This step is performed automatically (i.e., PSFRes=1) by DOLPHOT unless the use of residual PSF images is turned off (i.e., PSFRes=0).  These adjustments are made independently in each DOLPHOT science image.  The amplitude of the PSF adjustments provides a means to quantify the systematic uncertainty floor on the photometry for a given set of PSF models.  In \S \ref{sec:psf}, we provide the typical PSF adjustments made by DOLPHOT on ERS data relative to the default WebbPSF models.

Likewise, aperture corrections are normally computed automatically (i.e., ApCor=1), unless they are turned off in DOLPHOT (i.e., ApCor=0).  This calculation will estimate the magnitude difference between instrumental PSF-fitted magnitudes and magnitudes within a standard 10-pixel radius, accounting for the WebbPSF-predicted encircled energy within that standard radius.

Finally, zero points are applied to convert from instrumental magnitudes (in DN/sec) to the VEGAMAG system.    These zero points are in units of Jy for Vega (though Sirus is now the standard reference star) so they require conversion back from DN/sec to Jy using the FITS header keywords PHOTMJSR and PIXAR\_SR before the calibration is applied.  Zero points for NIRISS were provided relative to 1 DN/sec.  Thus, they are applied directly without additional conversion. Finally, we note that DOLPHOT parameters NIRCAMvega and NIRISSvega can be set to zero to report in ABmag instead of VEGAMAG.

\subsection{Post-Processing \& Catalog Creation}
\label{sec:cull}

Once complete, DOLPHOT saves all photometry data to a single ASCII file containing overall fit metrics (positions of objects in the coordinate system of the reference image, $\chi^2$, S/N, sharpness, roundness, crowding, and the object type: single pixel, point source, or extended source) for all objects identified.  Photometry and the same quality assessment information is also provided for all combined exposures in each filter (e.g., all F090W images), as well as for all individual exposures (e.g., each F090W images, which can be used for time domain studies, for example). The photometric data provided by filter and image include counts (DN), background, calibrated magnitude, and calibrated count rates, which can be useful if an upper limit is informative, such as in multi-wavelength SED fitting and time-domain studies \citep[e.g.,][]{gordon2016}.

As the DOLPHOT output includes all sources identified on the images, it is necessary to establish criteria for good detections (i.e., stars).   As part of this ERS program, \citet{warfield2023} developed a set of criteria to identify stars using DOLPHOT reported quality parameters. We adopt this scheme for this paper and classify good stars as those that satisfy all of the following criteria:

\begin{itemize}
\item $SNR_{F090W} \ge 5$
    \item $SNR_{F150W} \ge 5$
    \item $sharp_{F090W}^2 \le 0.01$
    \item $sharp_{F150W}^2 \le 0.01$
    \item $crowd_{F090W} \le 0.5$ 
    \item $crowd_{F150W} \le 0.5$ 
    \item $flag_{F090W} \le 3$
    \item $flag_{F150W} \le 3$
    \item{Object Type $\le 2$}
\end{itemize}

The sharpness parameter is zero for a perfectly-fit star, positive for a star that is too sharp (i.e., the flux is concentrated in a small number of pixels, e.g., a cosmic ray), and negative for a star that is too broad (perhaps a blend, cluster, or galaxy).   Our choice of SNR$\ge5$ is lower than the SNR$\ge10$ threshold in \citet{warfield2023}.  \citet{warfield2023} focused on optimizing the other parameters for star-galaxy separation, and therefore adopted a more conservative SNR threshold. 

The crowding parameter is in magnitudes. It reports how much brighter the star would have been measured had nearby stars not been fit simultaneously. For an isolated star, the value is zero. High crowding values are generally a sign of poorly-measured stars.

Error flags are defined as: 0 is a star that is recovered extremely well; 1 is that the photometry aperture extends off chip; 2 is that there are too many bad or saturated pixels; 4 is the center of the star is saturated; 8 is an extreme case of one of the above.  The DOLPHOT manual suggests using values of 3 or less in general or 2 or less for precision photometry. 

Object types are as follows: 1 is a good star; 2 is a star too faint for PSF determination; 3 is an elongated object; 4 is an object that is too sharp; 5 is an extended object.  As recommended by the DOLPHOT manual, we only keep object types 1 and 2 in our stellar catalogs.

As the deepest and highest angular resolution images, we found that applying these criteria to F090W and F150W photometry had the largest impact on the catalog culling \citep[e.g.,][]{warfield2023}.  For targeted science (e.g., luminous red stars at longer wavelengths) other criteria and/or application to other filters may produce more desirable results.  Similarly, cuts on single bands may be useful for particular science cases beyond CMDs (e.g., stellar SED fitting; \citealt{gordon2016}).  Finally, as discussed by \citet{warfield2023}, these criteria were focused on purity rather than completeness.  This is motivated by the large number of background galaxies present in our ERS imaging.   Less stringent cuts, particularly in sharpness, can produce deeper CMDs with less conservative completeness limits, albeit with a larger degree of non-stellar contamination.  We illustrate the effects of our fiducial culling criteria on our ERS targets in \S \ref{sec:ers_photometry}.  Readers who wish to explore alternative culling criteria can download our catalogs from the MAST high levels science products page.

\begin{table*}[]
\centering
\begin{tabular}{cccccc}
\toprule
Filter & $N_{exp}$ & Central Pixel & Central Pixel & Central Pixel & Photometric Error \\
 & & Model Mean & Model $\sigma$ & Mean $\delta$ &   \\
 & & (\% flux) & (\% flux) & (\% flux) & (mag) \\
 (1) & (2) & (3) & (4) & (5) & (6)\\
\toprule
NIRCAM F090W  & $88$ & $24.18$ & $3.98$ & $-0.69$ & $0.005$ \\
NIRCAM F150W  & $88$ & $16.63$ & $1.41$ & $-0.25$ & $0.001$ \\
NIRCAM F250M  & $8$  & $20.19$ & $2.07$ & $-0.24$ & $0.001$ \\
NIRCAM F277W  & $6$  & $18.41$ & $1.68$ & $-1.0$ & $0.005$ \\
NIRCAM F360M* & $8$  & $12.98$  & $0.76$ & $-1.57$ & $0.008$ \\
NIRCAM F430M  & $8$  & $10.22$ & $0.44$ & $0.07$ & $0.000$ \\
NIRCAM F444W  & $6$  & $10.05$ & $0.43$ & $-0.34$ & $0.002$ \\
NIRCAM F480M* & $8$  & $8.48$ & $0.30$ & $-0.41$ & $0.002$\\
NIRISS F090W  & $7$  & $31.47$ & $5.57$ & $-2.31$ & $0.014$ \\
NIRISS F150W  & $7$  & $28.53$ & $4.48$ & $-1.69$ & $0.010$ \\
\hline 
\toprule
\end{tabular}
\caption{Central pixel PSF data for the input WebbPSF PSF models (mean and standard deviation) and mean adjustments as measured by DOLPHOT from application to all ERS NIRCam and NIRISS images, except the 3rd exposure for M92.  Values in columns 3-5 are fractions of the total stellar flux.  Column 6 shows approximate photometric error in magnitudes caused by the application of the same PSF residual to all stars (see Equation \ref{eq:mag_err}).  Asterisks for NIRCAM F360M and F480M denote the filters only observed in the very sparse Draco II field; results from those filters have correspondingly higher uncertainties.}
\label{tab:psf_adjustments}
\end{table*}

\subsection{Point Spread Function Models}
\label{sec:psf}

The currently available NIRCAM and NIRISS modules incorporate model PSF libraries calculated using WebbPSF.  For this paper, the PSF models were generated using WebbPSF version 1.2.1, which adopts ``in-flight'' optical performance data (as opposed to pre-launch data). The alignment and stability of \jwst, including the wavefront and PSF, are known to vary over time \citep[e.g.,][]{McElwain2023}, which has the potential to affect photometry.  WebbPSF incorporates time-dependent optical path delay (OPD) maps to capture changes to the wavefront and PSF over time, enabling corrections for temporal changes to \jwst.

The WebbPSF model PSF library we implement in DOLPHOT consists of distorted PSF models for all the available NIRCam/NIRISS filters, oversampled by a factor of 5, calculated over a $51\times51$ physical pixel region. The models are calculated on a $5\times5$ spatial grid for each of the detector chips. The models were generated using OPD maps from July 24th 2022 (O2022072401-NRCA3\_FP1-1.fits). We used a G5V source spectrum from the Phoenix stellar library \citep[e.g.,][]{husser2013} to generate the PSF, which was sampled at 21, 9, and 5 wavelengths for wide, medium, and narrow bands, respectively.  v1.2.1 of WebbPSF includes the effects of charge diffusion and interpixel capacitance, which were not incorporated into previous WebbPSF models.  We found that the inclusion of these effects dramatically improves the quality of DOLPHOT photometry, including reducing photometric systematics by nearly an order-of-magnitude relative to WebbPSF models without these effects.  We discuss the the total photometric error budget in \S \ref{sec:error_budget}.

As described in \S \ref{sec:dolphot_photometry}, DOLPHOT makes adjustments to the model PSFs to provide improved matches to the data.  We summarize the effect of these PSF adjustments on the photometry in Table \ref{tab:psf_adjustments}.  This table provides the mean fractional central pixel brightness in each filter (i.e., the fraction of total PSF light in the central pixel averaged across all PSF models), the scatter in the central pixel fraction flux, and the mean PSF adjustment in the central pixel measured by DOLPHOT.  We computed these quantities across NIRCam and NIRISS images from our ERS program, except for the 3rd exposure of M92.

As DOLPHOT provides an average PSF correction (i.e., by computing the PSF adjustments on a set of bright, high SNR stars in each science image), we can quantify the uncertainty in this correction. Table \ref{tab:psf_adjustments} provides a simple estimate of the 1-$\sigma$ photometric error created by application of the same PSF residual image to all stars in a given science image.  The PSF adjustments are computed separately for each science image for each target.   That is, DOLPHOT has improved the model PSFs by adjusting the central pixel on average.  This is an improvement over using the native WebbPSF models without any adjustment, but it does still yield an uncertainty floor.  Considering the central pixel only, as DOLPHOT does for its PSF correction (see \citealt{dolphin2000b}), the magnitude error can be estimated as

\begin{equation}
    \sigma = \frac{2.5\,\sigma_{\rm model} \, \delta_{\rm mean}}{\log(10) \, \mu_{\rm model}^2}
    \label{eq:mag_err}
\end{equation}

\noindent where $\mu_{\rm model}$ is the mean model flux in the central pixel across all model PSFs in a given filter, $\sigma_{\rm model}$ is the scatter in the central pixel fluxes across all model PSFs in a given filter, and $\delta_{\rm mean}$ is the mean flux adjustment made by DOLPHOT. While this estimate is reasonably accurate for highly concentrated PSFs, such as the two NIRISS filters in our data sample, it becomes increasingly conservative for broader PSFs.

As Table \ref{tab:psf_adjustments} shows, the typical PSF adjustments, and the corresponding photometric uncertainties are small.  For all NIRCam filters, a modest amount of light is concentrated in the central pixel and the PSF adjustments are only typically a small fraction of the total flux.  The resulting photometric uncertainties introduced by the PSF models range from 0.001 to 0.008~mag.  For NIRISS, the PSFs are slightly sharper, i.e., more of the total light is centrally concentrated in the model PSFs, which leads to larger PSF adjustments. The corresponding photometric uncertainties are $\sim0.01$~mag in the NIRISS F090W and F150W filters.  Overall, the small adjustments needed by DOLPHOT to improve the WebbPSF models indicate that the models themselves are already quite good.  This was not the case with PSF models from previous versions of WebbPSF, all of which were systematically too sharp and the corresponding photometric errors, i.e., systematics from the PSF models, were at times larger than the photon counting noise.

We note that the values listed in Table \ref{tab:psf_adjustments} were computed for SW and LW images independently. We found all PSF adjustments to be marginally larger when running SW and LW images simultaneously, though the qualitative finding that the PSF adjustments are quite small (i.e., $\lesssim0.01$~mag) still holds.

\subsection{Artificial Star Tests}
\label{sec:ASTs}
While DOLPHOT provides an estimate of photometric uncertainties based on the goodness of fit and the noise characteristics of the data, a much better characterization of the photometric measurement uncertainties and selection function are accomplished through artificial star tests (ASTs). This long-established approach \citep[e.g.,][]{stetson1987, stetson1988} represents the ``gold standard'' in the field of resolved stellar population photometry. It relies on the injection of mock stellar sources into the raw images, which are then recovered using the identical photomeric procedure used to construct the raw and stellar DOLPHOT catalogs. The output of such simulated data tests can be used to quantify a number of aspects of data quality.  A common example is that the comparison between the input and output magnitudes of the mock catalog, as well as the fraction of mock stars that are successfully detected, provides a self-consistent characterization of photometric errors, systematic uncertainties, and photometric completeness as a function of spatial position in the images and location on the CMD. Throughout this paper, we focus on ASTs run only on the SW data, which illustrate the main points of ASTs.  The same procedures we describe can be used to run ASTs in an arbitrary number of bands, though the computational time can become quite expensive for large numbers of photometric bands.

The first step in running ASTs is to create a suitable input star catalog. For each target and camera, we created a list of $\gtrsim4\times10^5$ mock stars (see Table \ref{tab:ers_asts}), with positions drawn from uniform spatial distributions on the NIRCam/NIRISS footprints, aside from gaps between the chips and modules. For each star, we assign input magnitudes such that they are uniformly distributed over the F090W vs. (F090W-F150W) CMD with $17 \le$ F090W $\le 31$ and $-0.5 \le$ F090W$-$F150W $\le 2$.  

The artificial stars are then injected into all science (i.e., \texttt{cal}) images using the best PSF model and realistic noise obtained from the original reduction run. The stars are injected one at the time, to avoid altering the crowding properties of the original images. The star magnitude and position is then measured by DOLPHOT, as if it were a real source. Performing this operation for all the input stars, we obtain a catalog of output magnitudes, positions, and goodness-of-fit parameters. This list is then culled using the same quality criteria applied to the original photometric catalog (cf. \S~\ref{sec:cull}).

\subsection{Optimizing Photometric Parameters}
\label{sec:optimizing}

DOLPHOT is a flexible code in that it provides the user extensive control over details of the photometric reduction (e.g., PSF adjustments, sky fitting, image alignment methods, noise models).  This ensures that DOLPHOT can produce high-quality photometry for a wide variety of images (e.g., crowded vs.\ uncrowded, presence/absence of surface brightness gradients, images from multiple filters with widely different characteristics).  At the same time, this flexibility is characterized by a large number of parameters that can require tuning in order to produce ``optimal'' photometry.  DOLPHOT parameters have been refined in the context of major \hst\ programs over the past decade, culminating in a set of parameters recommended by the PHAT program that encompass a wide range of image properties \citep[e.g.,][]{williams2014}.  While these parameters produce excellent \hst\ photometry, it is important that we investigate parameters that may provide better photometry for \jwst.

Accordingly, we performed a large set of DOLPHOT runs to explore the effect of changing a select set of DOLPHOT parameters. We performed these runs on all three ERS targets, to experiment with different stellar density regimes: high crowding for WLM, low crowding for M92, and an almost empty field for Draco II. As the SW and LW channels have different detector characteristics (e.g., plate scale), we executed the experiments on the two channels independently. For each target and each channel, we explored different values of \texttt{FitSky} (which sets the method for local sky measurement), \texttt{RAper} (which sets the size of the aperture in which photometry is performed), and \texttt{Rchi} (which sets the size of the region over which the fit is evaluated). We set the value of \texttt{FitSky} to either 2 (fit the sky inside the PSF region but outside the photometry aperture) or 3 (fit the sky within the photometry aperture as a 2-parameter PSF fit). For a given \texttt{FitSky}, we let \texttt{RAper} range between 1 and 5 pixels, in discrete increments, and \texttt{Rchi} range between 0.5 and \texttt{RAper}, in increments of 0.5 pixels. In the \texttt{FitSky}=2 case, the values of \texttt{Rsky2} (which set the inner and outer radius for sky computation) are also adjusted and set to \{\texttt{RAper}+1; 2.5(\texttt{RAper}+1)\}. For \texttt{FitSky}=3, we explored an additional grid, defined by \texttt{RAper} values of 7, 10, and 13 and \texttt{Rchi} values of 1.5, 2, and 3. This exploration results in 69 parameter permutations per channel, per target, totaling 414 DOLPHOT runs, which consumed nearly five years of CPU time.

For each of these runs, we used thousands of ASTs (see \S~\ref{sec:ASTs}) as one metric for evaluating photometric performance.  ASTs were injected at various locations on the CMD (from very bright, high-SNR parts down to the detection limit). We then used these stars to quantify the mean scatter in color and magnitude, and the completeness fraction at each CMD location. Inspection of these metrics revealed that in large regions of the parameter space, DOLPHOT performance was poor. For example, most runs  
 with $\texttt{Rchi}>3.0$ and/or  \texttt{FitSky}=3 produced obviously poor photometry (e.g., low number of stars, poor completeness, large scatter). Adopting \texttt{Fitsky}=2, we identified a small region of the \texttt{RAper}-\texttt{Rchi} parameter space ($2\leq\texttt{RAper}\leq4$ and $1.5\leq\texttt{Rchi}\leq2.5$) where DOLPHOT provided the ``best'' photometry. Within this parameter region, the photometric performance was fairly comparable. Different permutations of these parameters produced slight trade-offs in completeness versus photometric precision, though the differences were generally at the few percent level or less. We also found a slight trend with crowding, in that the optimal \texttt{RAper} value would increase as the field became less crowded, but again, this effect was small.

 Given the similarity of the photometry over this parameter space, we decided to adopt a single set of parameters for each instrument and channel.  Part of the motivation behind this choice is to provide the community with easy-to-use guidance for DOLPHOT reductions that also produces high-quality photometry.   The recommended PHAT parameters all live within the ``optimal'' DOLPHOT parameters we have identified for \jwst.  Therefore, we decided to adopt the PHAT set up as our \jwst\ DOLPHOT parameters.   Specifically, we recommend the PHAT WFC3/IR parameters for the NIRCam SW channel and the ACS/WFC parameters for the NIRCam LW channel and NIRISS. The full parameter set is listed in Table~\ref{tab:setup}.

\begin{figure*}[ht!]

\epsscale{1.2}
\plotone{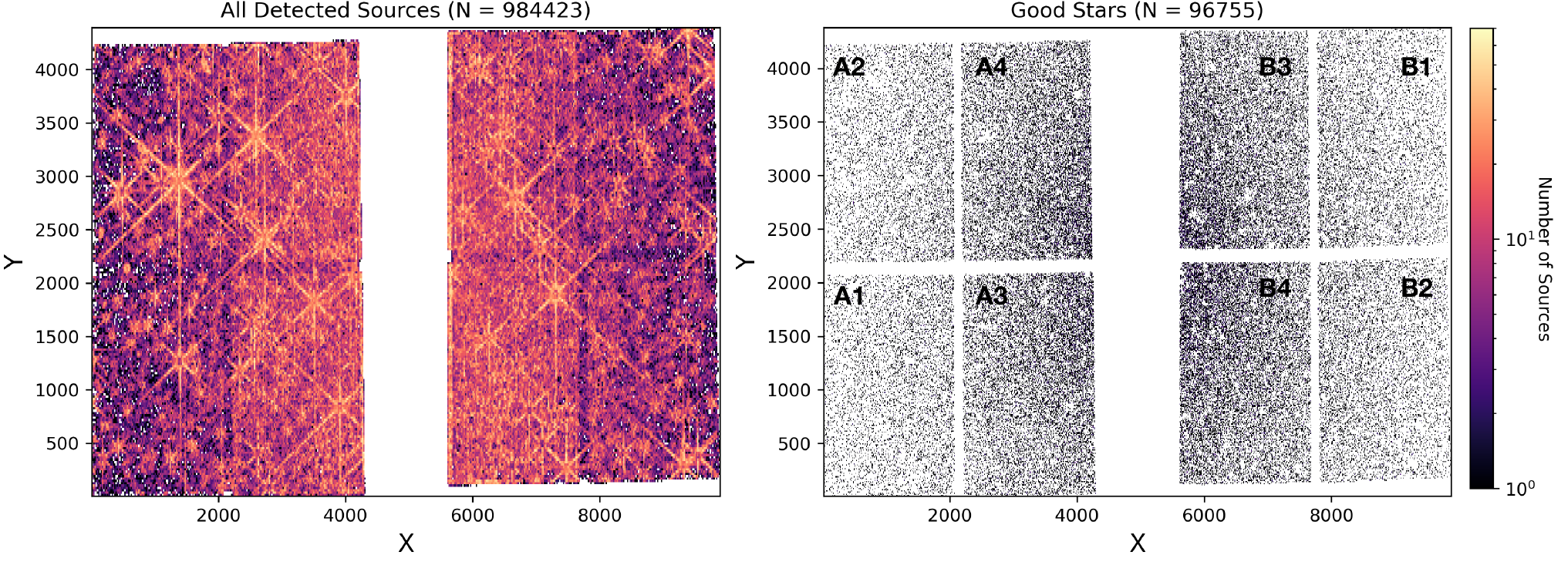}
\caption{The spatial distribution of objects and stars detected by DOLPHOT in our NIRCam imaging of M92, excluding the 3rd exposure.   \textbf{Left:} A density map ($\sim50\times50$ pixel bins) of the $\sim 9.8\times10^5$ objects reported in the raw DOLPHOT catalog. There is a density gradient toward the center of the image and cluster center. Additional features include a high density of sources that trace saturated stars and a lower density of sources in the SW chip gaps.  \textbf{Right:} The density map of stars that passed the catalog culling criteria listed in \S~\ref{sec:cull}.  These criteria removed the vast majority of obvious artifacts (e.g., corresponding to saturated stars, diffraction spikes) and reveal a clear stellar density gradient as is expected for a GC. Note that the individual SW chips are labeled.}
\label{fig:m92_spatial}
\end{figure*}

\begin{table}[]
\centering
\begin{tabular}{clcc}
\toprule
Target & \multicolumn{1}{c}{Camera} & Object Density                   & Stellar Density                  \\
       & \multicolumn{1}{c}{}       & ($N$ / arcsec$^2$) & ($N$ / arcsec$^2$) \\
(1)                          & \multicolumn{1}{c}{(2)} & (3)  & (4)  \\
\toprule
\multicolumn{1}{l}{M92}      & NIRCam                  & 28.1 & 2.8  \\
                             & NIRISS                  & 7.4  & 0.2  \\
\multicolumn{1}{l}{WLM}      & NIRCAM                  & 48.8 & 13.2 \\
                             & NIRISS                  & 8.9  & 0.7  \\
\multicolumn{1}{l}{Draco II} & NIRCAM                  & 14.7 & 0.03 \\
\toprule
\end{tabular}
\caption{The average densities of objects (Column 3) and stars (Column 4) from the DOLPHOT photometric reductions of our ERS fields.}
\label{tab:ers_density}
\end{table}

\begin{figure*}
\gridline{\fig{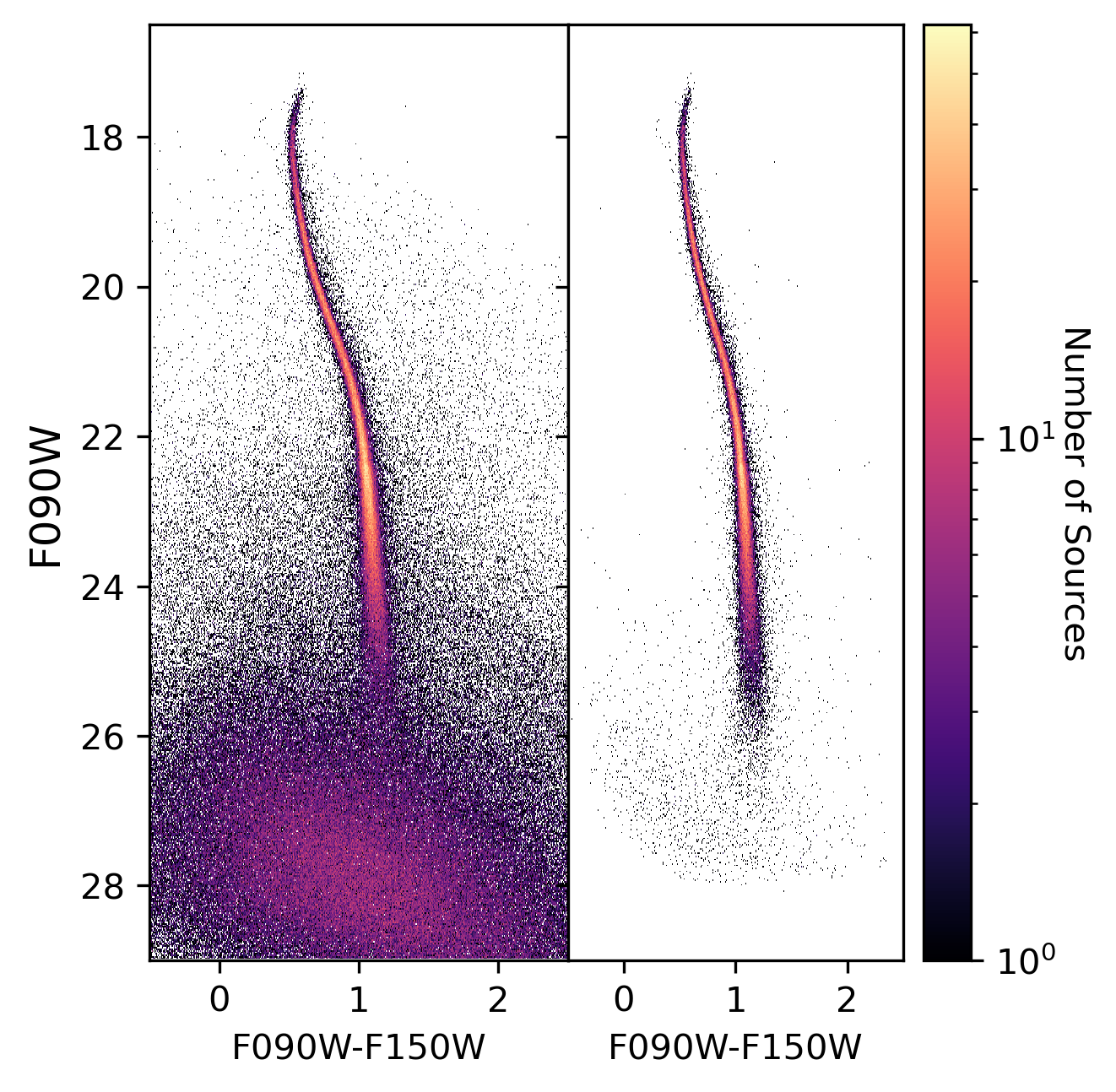}{0.34\textwidth}{(a)}
\fig{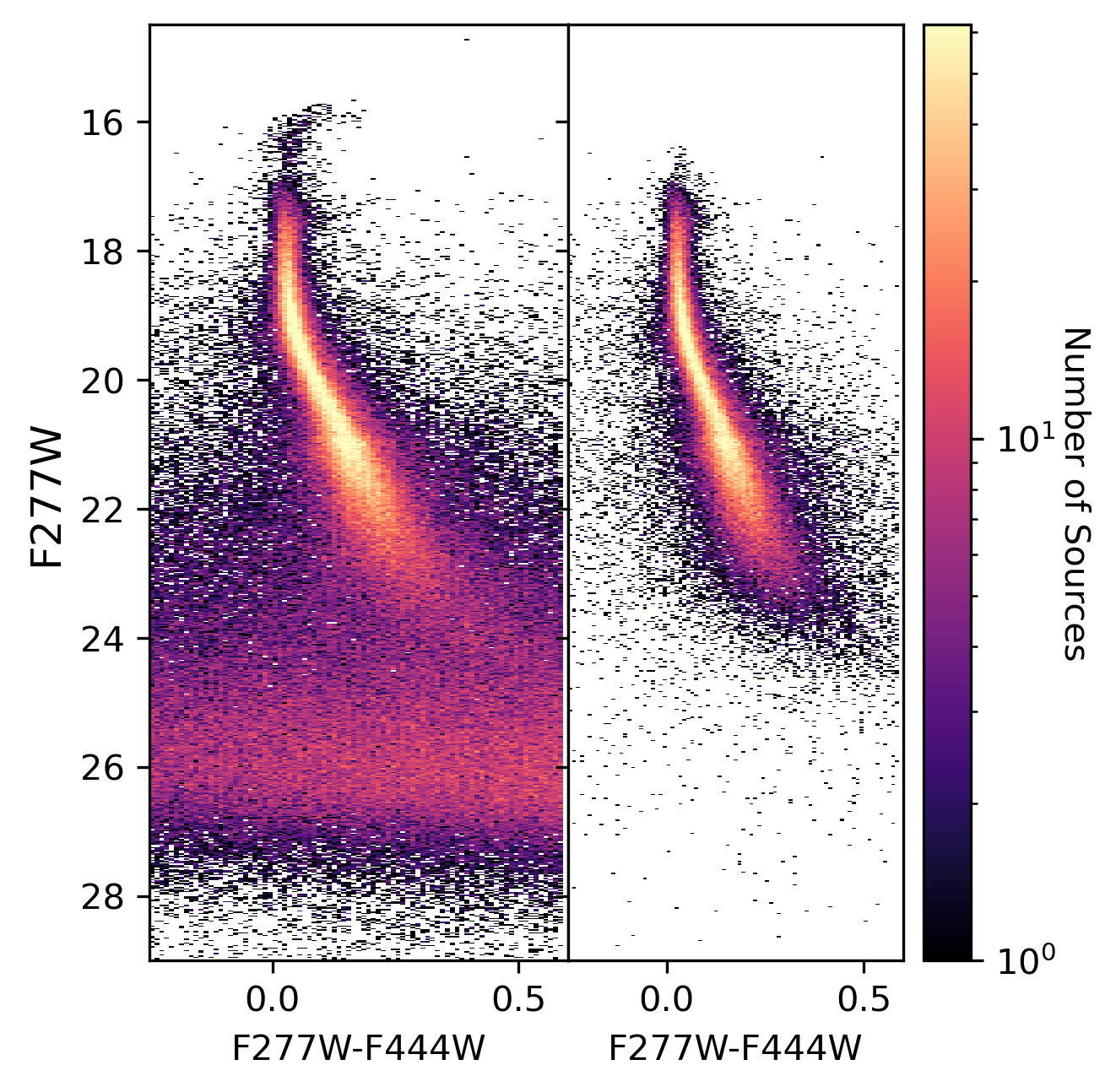}{0.34\textwidth}{(b)}
\fig{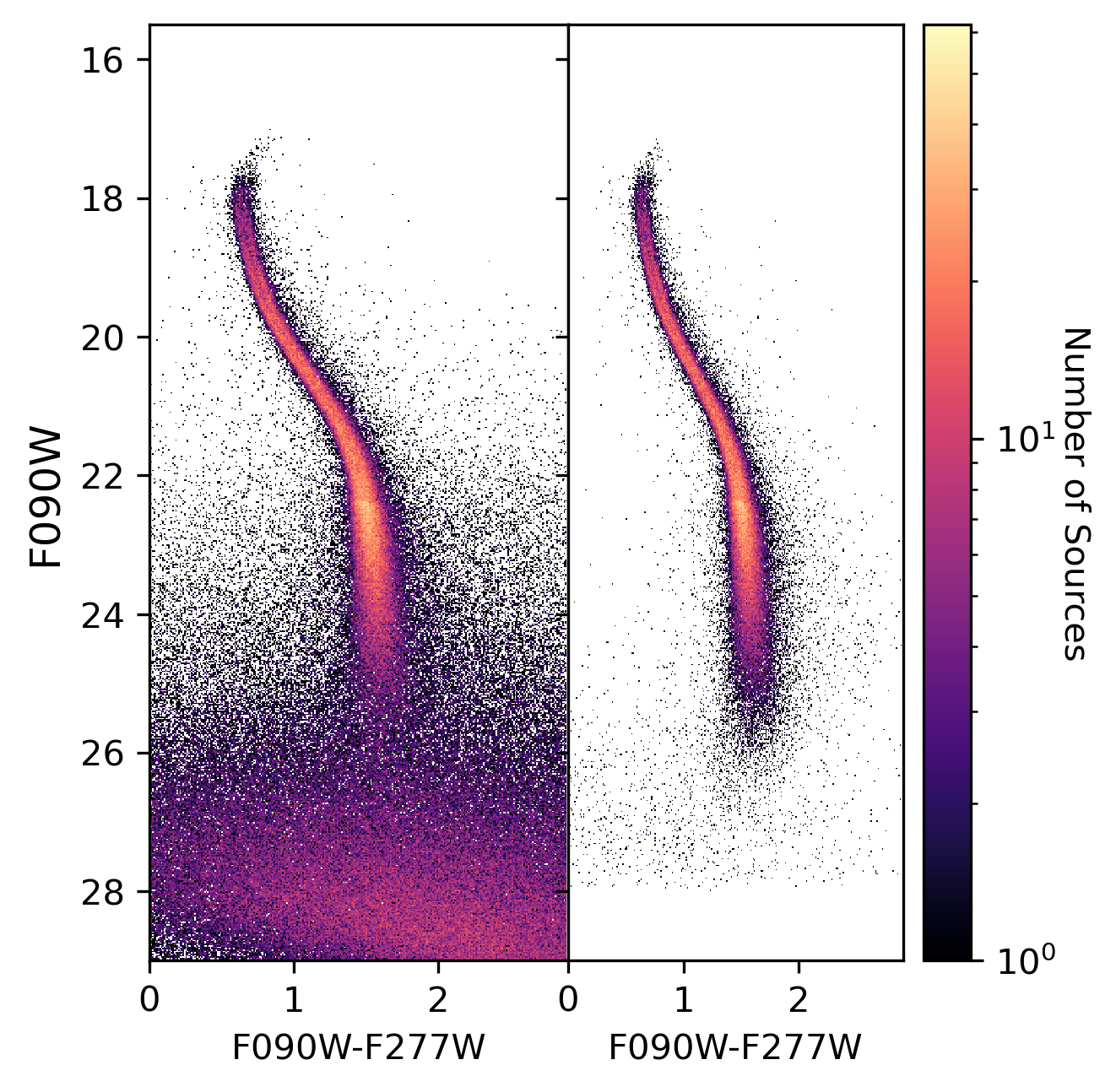}{0.34\textwidth}{(c)}}
\caption{NIRCam CMDs of M92 in select filter combinations.  These CMDs exclude the 3rd exposure.   Panels (a), (b), and (c) show the CMDs for the SW (F090W$-$F150W), LW (F277W$-$F444W), and an example SW/LW (F090W$-$F277W) filter combination.  The left side of each plot shows CMDs of all objects detected, while the right panels are CMDs of stars that passed the culling criteria.  In all cases, the photometry is of excellent quality and the culling critria removes a large fraction of obvious non-stellar sources.  There appear to be two MSs in the F090W$-$F277W CMD that are offset in color from one another.  This is due to small zero point differences between the NIRCam chips/modules.
\label{fig:m92_cmds}}
\end{figure*}

\begin{figure}[ht!]
\plotone{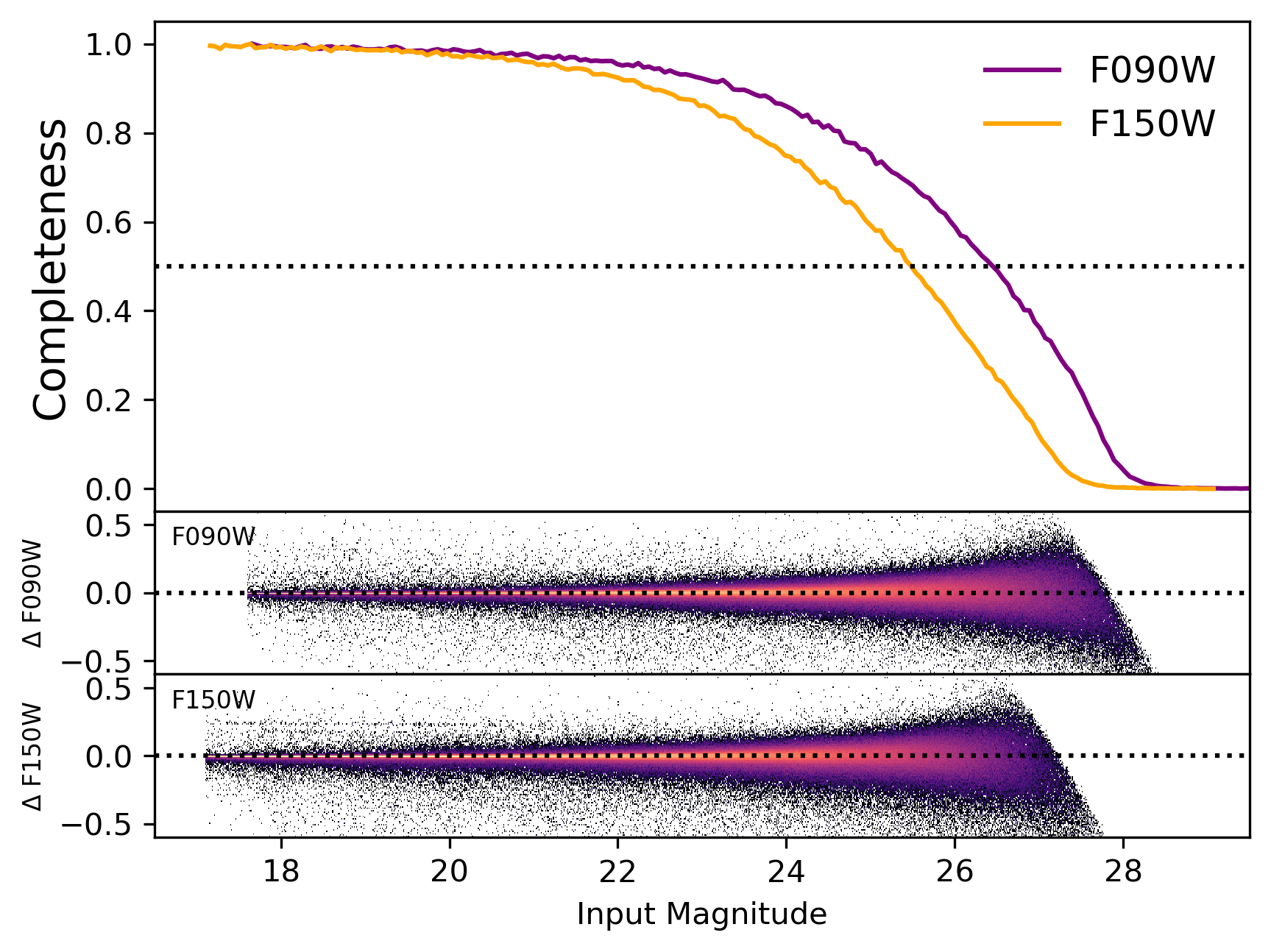}
\caption{The completeness and photometric uncertainty for the M92 NIRCam F090W and F150W data computed from artificial star tests.  The top panel shows the completeness function for each filter.  The bottom panels show the recovered minus the input magnitude difference for ASTs that pass the same culling criteria as applied to the photometry.  The 50\% completeness limits are $m_{F090W}=26.4$ and $m_{F150W}=25.4$.  Both filters have a bias consistent with zero. \label{fig:m92_asts}}
\end{figure}

\begin{figure*}[ht!]
\epsscale{1.2}
\plotone{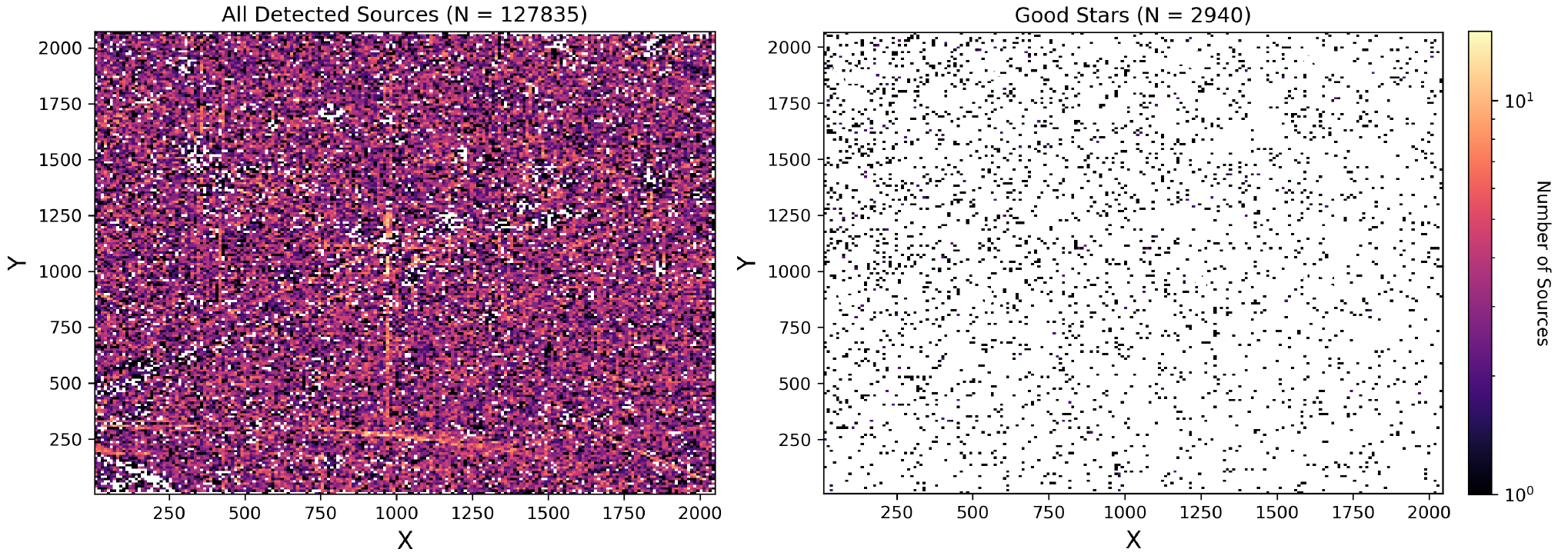}
\caption{The spatial distribution of objects and stars detected by DOLPHOT in our NIRISS imaging of M92.   \textbf{Left:} A density map ($\sim50\times50$ pixel bins) of the $\sim 1.3\times10^5$ objects reported in the raw DOLPHOT catalog.  The spatial distribution of detected objects is uniform.  Some artifacts (e.g., bright stars, claws; \citealt{rigby2022}) are visible. \textbf{Right:} The density map of the $\sim2.9\times10^3$ stars that passed the catalog culling criteria listed in \S \ref{sec:cull}.  These criteria removed the vast majority of obvious artifacts (e.g., corresponding to saturated stars, diffraction spikes), leaving a sparse sampling of stars.  This low density is expected, given that the field is located at $\sim$ 5 half-light radii.}
\label{fig:m92_spatial_niriss}
\end{figure*}

\begin{figure}
    \centering
\plotone{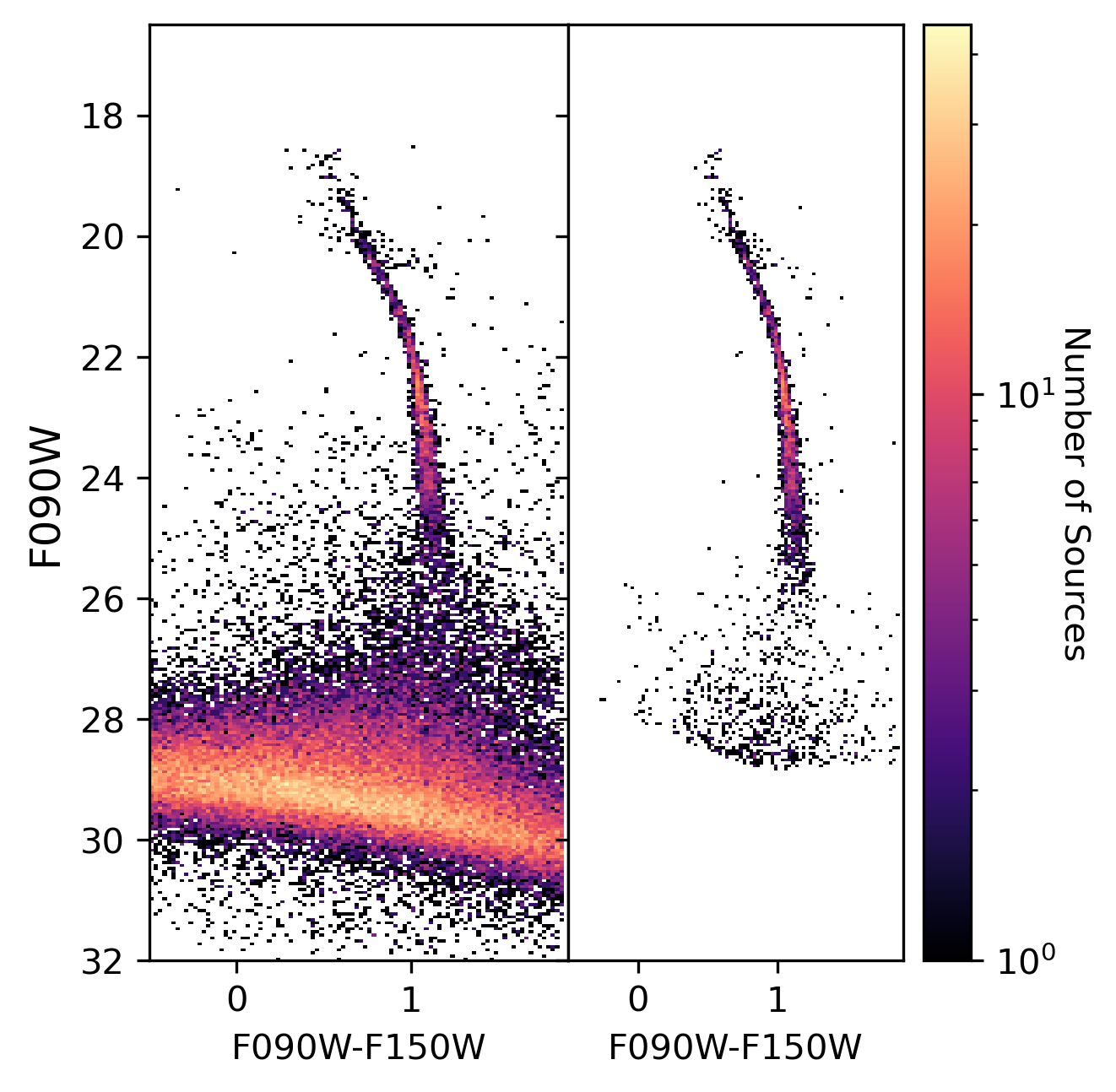}
    \caption{The NIRISS CMD of M92, excluding the 3rd exposure.  The left panel shows all objects detected, the right panel shows the stars that passed the culling criteria.  Compared to the SW NIRCam CMD in Figure~\ref{fig:m92_cmds}, the NIRISS CMDs cover a smaller dynamic range in luminosity, owing to saturation effects at the bright end and lower sensitivity at the faint end.  The diagonal feature of stars at F090W$\sim20$ corresponds to objects in image artifacts that were not removed by the applied culling criteria. The increased scatter in this CMD, relative to the SW NIRCam CMD, is due to the combination of lower SNR as well as WebbPSF models that are not as well-matched to the observed PSFs.}
    \label{fig:m92_cmd_niriss}
\end{figure}

\begin{figure*}[t!]

\epsscale{1.2}
\plotone{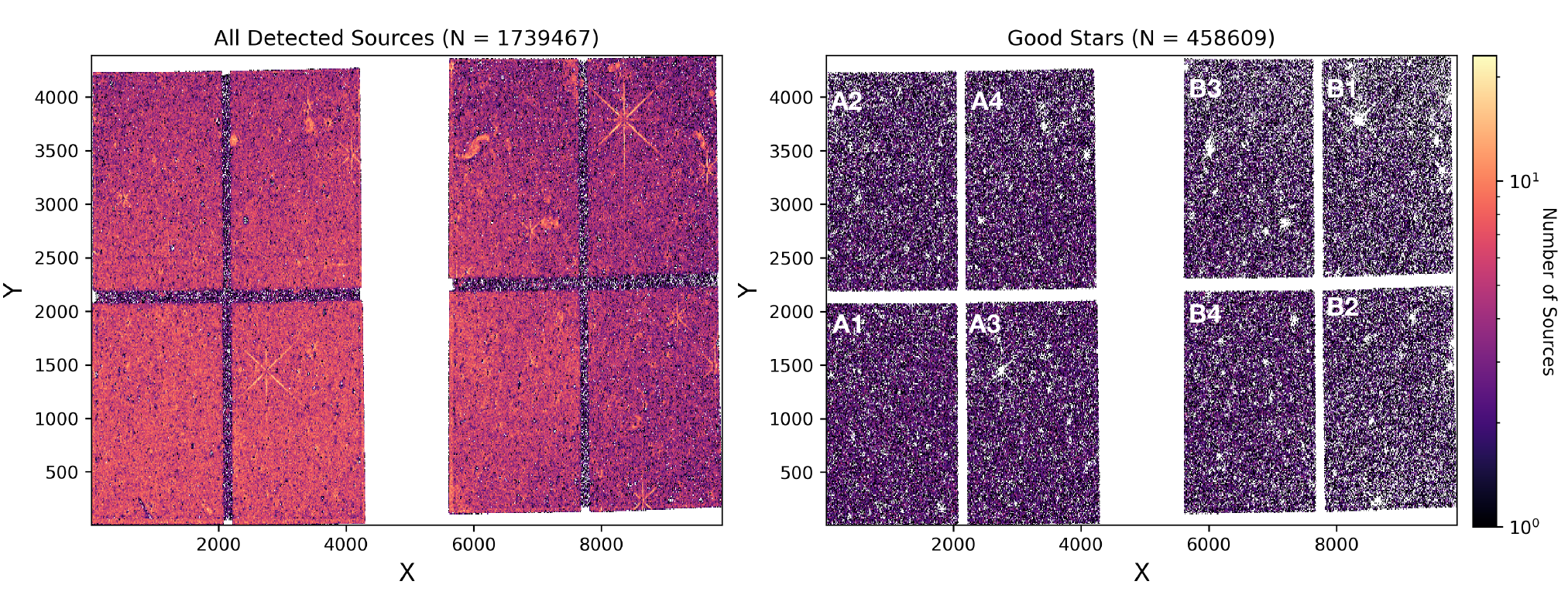}

\caption{The spatial distribution of objects and stars detected by DOLPHOT in our NIRCam imaging of WLM.   \textbf{Left:} A density map ($\sim50\times50$ pixel bins) of the $\sim 1.7\times10^6$ objects reported in the raw DOLPHOT catalog.  The spatial distribution of detected objects is fairly uniform, with a slight gradient toward chip A1, which is positioned closest to the center of the galaxy. Additional features include a high density of sources that trace saturated stars and a lower density of sources in the SW chip gaps.  \textbf{Right:} The density map of the $\sim4.6\times10^5$ stars that passed the catalog culling criteria listed in \S~\ref{sec:cull}.  These criteria removed the vast majority of obvious artifacts (e.g., corresponding to saturated stars, diffraction spikes). The density of stars increases toward chip A1, which is closest to the center of the galaxy.}
\label{fig:wlm_spatial}
\end{figure*}

\begin{table}[]
\centering
\begin{tabular}{ccc}
\toprule
Detector & Parameter & Value \\

\toprule
NIRCam/SW&RAper&2\\
NIRCam/SW&Rchi&1.5\\
NIRCam/SW&Rsky2&``3 10"\\
NIRCam/LW&RAper&3\\
NIRCam/LW&Rchi&2.0\\
NIRCam/LW&Rsky2&``4 10''\\
NIRISS&RAper&3\\
NIRISS&Rchi&2.0\\
NIRISS&Rsky2&``4 10''\\
All&FitSky&2\\
All&PSFPhotIt&2\\
All&PSFPhot&1\\
All&SkipSky&1\\
All&SkySig&2.25\\
All&SecondPass&5\\
All&SigFindMult&0.85\\
All&MaxIT&25\\
All&NoiseMult&0.1\\
All&FSat&0.999\\
All&FlagMask&4\\
All&ApCor&1\\
All&Force1&0\\
All&PosStep&0.25\\
All&RCombine&1.5\\
All&SigPSF&5.0\\
All&PSFres&1\\
All&InterPSFlib&1\\
All&UseWCS&2\\
All&CombineChi&0\\
\hline 
\toprule
\end{tabular}
\caption{Recommended DOLPHOT input parameters based for NIRCam SW, LW, and NIRISS imaging based on the extensive photometric testing described in \S \ref{sec:optimizing}. A detailed description of each parameter can be found in the DOLPHOT manual and on our ERS documentation webpage.}
\label{tab:setup}
\end{table}

\begin{figure*}[t!]
\gridline{\fig{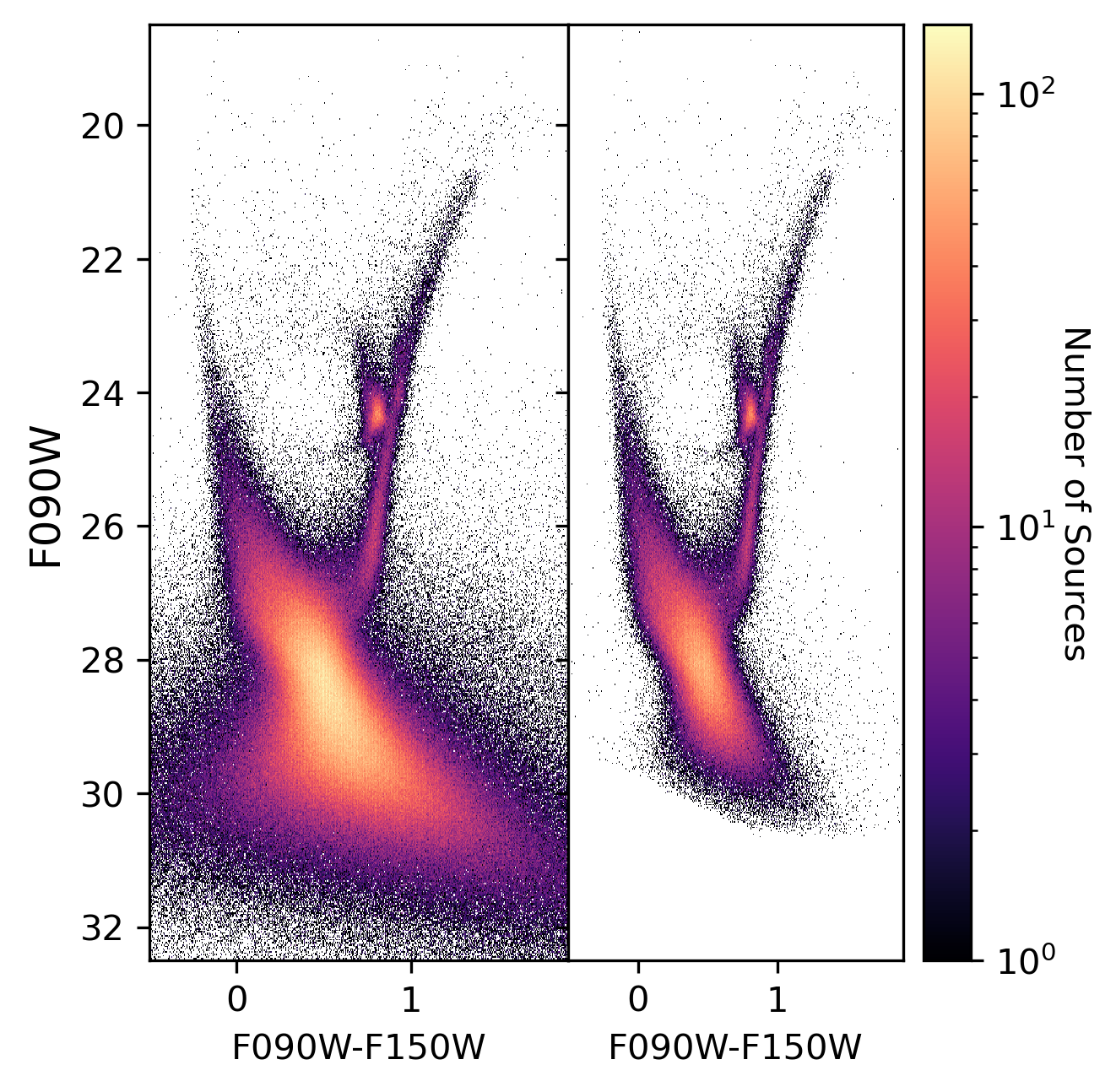}{0.34\textwidth}{(a)}
\fig{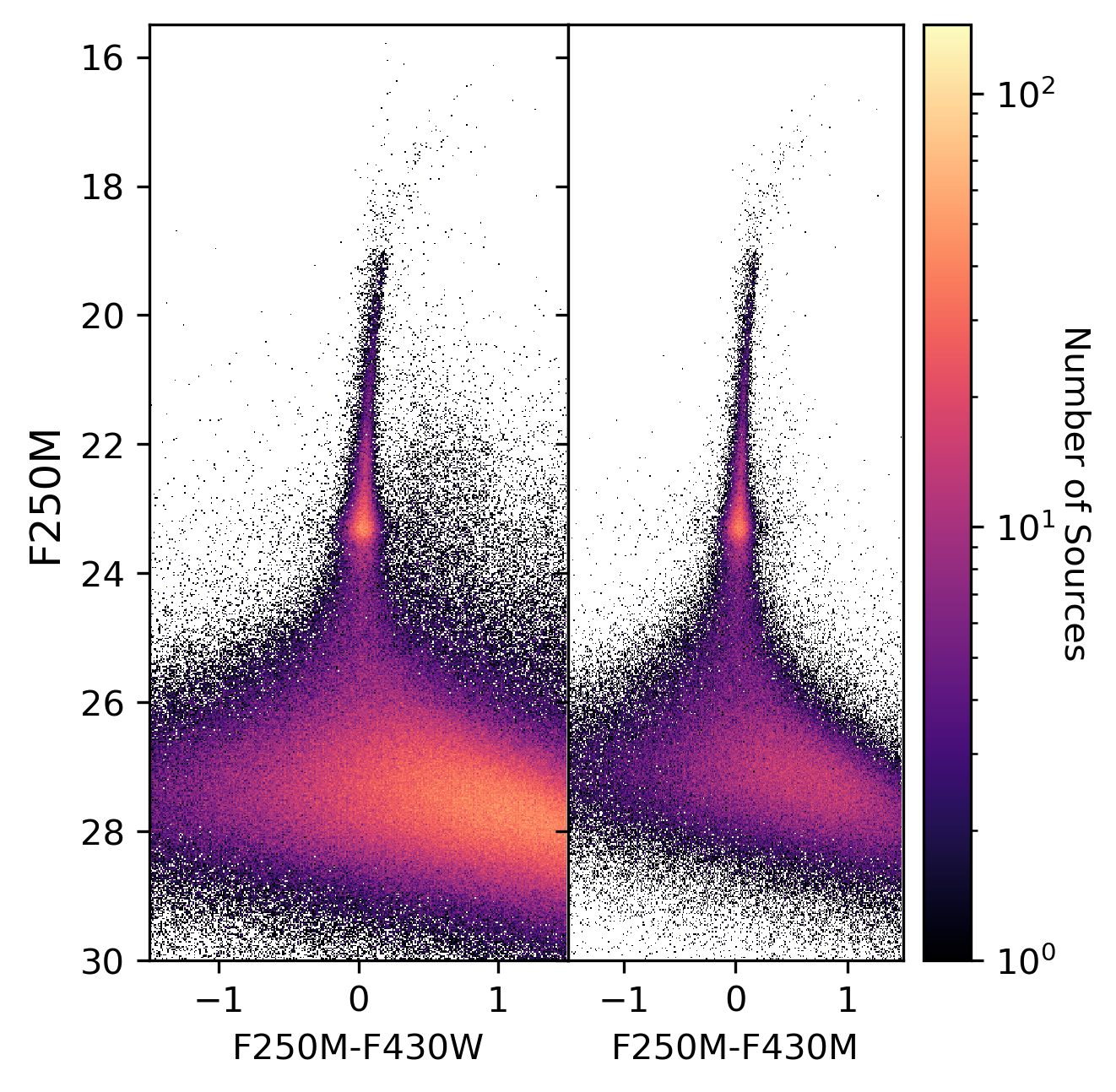}{0.34\textwidth}{(b)}
\fig{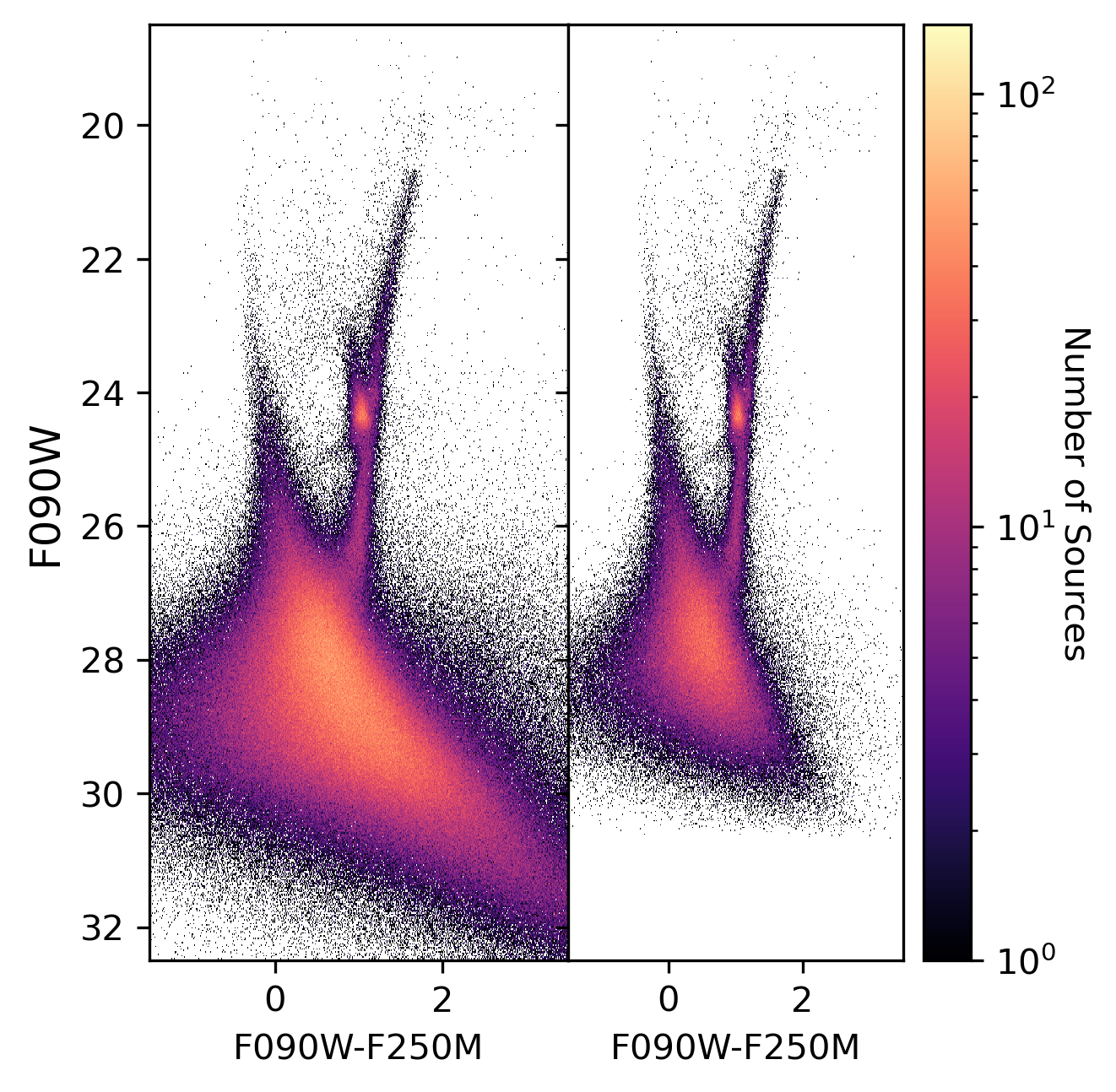}{0.34\textwidth}{(c)}}
\caption{NIRCam CMDs of WLM in select filter combinations.    Panels (a), (b), and (c) show the CMDs for the SW (F090W$-$F150W), LW (F250M$-$F430M), and an example SW/LW (F090W$-$F250M) filter combination.  The left side of each plot shows CMDs of all objects detected, while the right sides are CMDs of stars that passed the culling criteria described in \S~\ref{sec:cull}.  The SW CMD of WLM is the deepest ever constructed for a galaxy that is not within the viral radius of the MW.  A number of stellar evolution sequences are quite tight (e.g., RGB, sub-giant branch, young MS), in line with the exquisite SNR.   Panel (b) shows the LW-only CMD, which shows a bright AGB star sequence, a bright RGB, and a well-populated RC.  The increased scatter below the red clump is the result of culling criteria applied only to SW data.  Panel (c) shows an example SW-LW CMD.  As with panel (a), a number of clear sequences emerge.  The CMD nearly reaches the oldest MSTO, despite the LW filter being a medium band.   
\label{fig:wlm_nircam_cmd}}
\end{figure*}

\section{DOLPHOT Photometry of Early Release Science Data}
\label{sec:ers_photometry}

In this section, we present DOLPHOT photometric reductions of our ERS NIRCam and NIRISS data.  We used the procedures described in \S \ref{sec:dolphot} and the DOLPHOT parameters listed in Table~\ref{tab:setup} for all targets.  For each target and instrument, we discuss examples that illustrate the results of our DOLPHOT runs.  The full catalogs are available on MAST for interested readers to download.  Similarly, step-by-step details for our runs are available on our ERS documentation page.  Comparisons between our reductions and stellar models (i.e., to demonstrate the reasonability of the zero points and stellar models, which have not changed since our past publication) were already made in \citet{weisz2023a}, and we do not repeat those comparisons here.  We do not include the \dracoii\ NIRISS field in our analysis as there were not enough bright stars in the field for astrometric alignment in DOLPHOT.

\subsection{M92}

\subsubsection{NIRCam}

Figure \ref{fig:m92_spatial} shows the the spatial distribution of objects (left) and stars (right) recovered by our application of DOLPHOT to the M92 NIRCam imaging.  This M92 DOLPHOT reduction excludes the 3rd exposure for reasons discussed in \S \ref{sec:obs} and in Appendix \ref{sec:m92_3rd}.

In total, DOLPHOT finds $\sim9.8\times10^5$ sources in the NIRCam field.  This translates to a density of  $\sim28$ objects per arcsec$^2$ (Table \ref{tab:ers_density}). The spatial distribution of all objects (left panel) shows that a sizable number of these objects are obvious artifacts and not M92 stars.  Visually, the most obvious contaminants are sources associated with bright, saturated stars.  These objects trace both the cores and diffraction spikes.  Though less obvious visually, there are a large number of background galaxies in the spatial plot of all objects.  The SW interchip gaps are clearly visible in both spatial plots. All sources in these chip gap regions correspond to objects detected in LW filters only, as there is no SW coverage in the chip gaps owing to the different detector shapes and our choice not to fill the gaps by dithering. The gaps are completely empty in the star-only plot (right panel) because  
our nominal culling criteria require detections in the SW filters.

The right panel of Figure \ref{fig:m92_spatial} shows the spatial distribution of stars in the M92 NIRCam field (i.e., objects that passed the culling criteria defined in \S \ref{sec:cull}).  The culled catalog contains $9.7\times10^4$ stars, which is $\sim10\%$ of the total number of objects detected.  Visually, the spatial distribution of stars qualitatively follows what is expected of a globular cluster \citep[e.g.,][]{king1962}: a higher concentration of stars in the center, with a decrease in density as a function of increasing radius.  Many of the obvious artifacts have been removed such as objects associated with saturated stars and more extended background galaxies.  As discussed in \citet{warfield2023}, these culling criteria are designed with purity in mind, though they are not perfect, as some objects associated with diffraction spikes and compact background galaxies can still be mistaken for stars and included in the culled catalogs.  Though these only represent a small fraction of the \textit{bona fide} M92 stars, careful inspection is required if individual stars/objects are of interest (e.g., those that occupy sparsely populated regions of a CMD).

Figure \ref{fig:m92_cmds} shows an illustrative set of CMDs for all objects (left panels) and stars (right panels) in the M92 NIRCam field for select SW and LW filter combinations.  Panel (a) shows the F090W$-$F150W CMDs.  The effects of the culling criteria are quite dramatic, particularly at faint magnitudes.  The majority of non-stellar sources are located at the very bottom of the CMD (i.e., low S/N) or in regions of the CMD not typically occupied by stellar sources.  The application of the culling criteria removes $\sim90$\% of the detected objects, producing the exquisitely deep CMD shown in the right panel.  The resulting stellar CMD shows a very tight lower main sequence as expected for a metal-poor GC.  The bright end of the CMD begins at the main sequence turn off (MSTO), while fainter features such as the MS kink and bottom of the stellar sequence (i.e., \mstar$\sim0.1$ \msun) are evident.  These features are discussed in \citet{weisz2023a}.  The sparse collection of objects near the bottom of the SW CMD are some combination of compact background galaxies that were not picked up by the culling criteria and a small number of white dwarfs \citep{nardiello2022}.  Brown dwarfs are likely too faint to be included in this CMD \citep[e.g.,][]{dieball2019}.   We examined the CMDs as a function of SW chip and found them to be in generally good agreement. 

\begin{table}[]
\centering
\begin{tabular}{clccc}
\toprule
Target                       & \multicolumn{1}{c}{Camera} & $N_{\rm ASTs}$ & F090W & F150W \\ 
                       & \multicolumn{1}{c}{} & & 50\% Comp. &  50\% Comp. \\

                        & \multicolumn{1}{c}{}    &        & (mag) & (mag) \\ 
(1)                     & \multicolumn{1}{c}{(2)} & (3)    & (4)   & (5)   \\ 
\toprule
\multicolumn{1}{l}{M92} & NIRCam                  & 2871773 & 26.4  & 25.4 \\ 
                        & NIRISS                  & 1168289 & 28.1  & 27.4   \\
\multicolumn{1}{l}{WLM} & NIRCAM                  & 1573112 & 28.7  & 27.7 \\
                        & NIRISS                  & 1168465 & 29.3  & 28.5   \\ \multicolumn{1}{l}{Draco II} & NIRCAM                     & 408455         & 29.6             & 28.3      \\ 
\toprule
\end{tabular}
\caption{Summary of the SW artificial star tests for each of our ERS targets.  Column (3) lists the number of ASTs run; Columns (4) and (5) list the F090W or F150W magnitude corresponding to the 50\% completeness limit.}
\label{tab:ers_asts}
\end{table}

Panel (b) shows the LW NIRCam CMD of M92.  Application of the culling criteria, which is based only on the SW data, removes many artifacts and produces one of the deepest mid-IR CMDs of a GC to date.  The LW CMD includes the MSTO at the bright end, and extends only to the middle of the MS kink at the faint end.  Culling criteria tailored specifically to the LW channels may be able to produce a slightly deeper CMD with fewer points away from the MS.  However, a full exploration of filter dependent culling criteria is beyond the scope of this paper.  For interested readers, this exercise can be readily done with the public ERS catalogs we provide on MAST.

The LW CMD shows some structure at the brightest magnitudes of the LW near the MSTO.  The source is like related to the nuances of saturation and star locations with respect to the center of a pixel.  Savino et al.\ in prep. discusses these effects in more detail.

Panel (c) shows the F090W$-$F277W CMD of the M92 NIRCam field.  As with the other example CMDs, the culling criteria provide for the removal of many non-stellar sources.  The resulting CMD extends from the MSTO at the bright end to the bottom of the MS at the faint end.   A close inspection of this CMD shows a slight bifurcation in the MS, that is most visible near the MSTO.  The two MSs are offset by $\sim0.05$~mag.  Closer inspection of our data shows that the color of the sequences changes as a function of NIRCam chip/module.  The offsets are most apparent in the SW$-$LW CMDs (e.g., it is also clear in the F150W$-$F444W CMD), and are far smaller in the SW-only CMDs, and certainly not as large as our team initially reported in \citet{boyer2022}.  Our findings indicate that the spatial zero points need further refinement, which is a goal of the \jwst\ absolute flux calibration program \citep{gordon2021}.  Updates to the zero points should produce even tighter sequences in M92.

Figure~\ref{fig:m92_asts} shows the SW completeness functions (top panel) and photometric bias and scatter (bottom panels) as determined from $\sim10^6$ ASTs inserted into the NIRcam images of M92.  The shape of the completeness functions behaves as expected for a mostly uncrowded stellar field. The completeness is $>$50\% for the entirety of the stellar sequence, reaching 50\% at $m_{F090W}=26.4$ and $m_{F150W}=25.4$.  The completeness gradually decreases until it reaches zero at F090W$\sim28.1$ and F150W$\sim27.6$.  

The bottom two panels show the difference between the recovered and input magnitudes for ASTs that pass the culling criteria.   The mean of both distributions is 0, which indicates no bias in the AST recovery. The scatter increases as a function of magnitude in the expected manner for ASTs \citep[e.g.,][]{monelli2010a, dalcanton2012b}.  The LW filter for M92, and the other targets in our program, has similar AST characteristics -- no bias and scatter that behaves as expected.

Two other studies have previously published CMDs of M92 using \jwst\ ERS imaging.  \citet{nardiello2022} and \citet{ziliotto2023} performed photometry on our M92 NIRCam imaging using empirical PSFs, based on the method of \cite{anderson2000}.  In general, the CMDs produced by their programs and DOLPHOT are qualitatively similar, i.e., deep and high SNR.  

However, there are a number of subtle details in the reduction procedures that affect interpretation of the results at the few to several percent level of precision and accuracy. For example, \citet{ziliotto2023} use the 3rd exposure of M92, which adds extra noise (see Appendix \ref{sec:m92_3rd}) and do not note the module dependent zero point offset in F277W/F444W that was present prior to the zero point updates provided by STScI in Fall 2023. We found these to introduce non-negligible bias and scatter in the LW photometry. In principle, such scatter could enhance multiple population effects they identify.  

Similarly, the rapid publication timescale of \citet{nardiello2022} meant much of the calibration work on \jwst\ was incomplete or unavailable.  Their results preceded post-launch STScI zero points, stable filter curves, flat field updates, and usable DQ arrays.  They circumvented many of these issues by anchoring their photometry to theoretical predictions from the BaSTI stellar evolution models \citep{hidalgo2018}. This has the effect of producing qualitatively good-looking CMDs, but it also glosses over many of the calibration effects for which we chose M92 as an ERS target. Our program was designed to obtain high S/N of a target with a simple stellar population in order to help diagnose potential shortcomings, systematics, etc., as we have done in this paper.

\subsubsection{NIRISS}

Figure~\ref{fig:m92_spatial_niriss} shows the spatial distribution of $1.3\times10^5$ objects (left) and $2.9\times10^3$ stars (right) for our NIRISS M92 field.  The objects are essentially uniformly distributed across the field.  There are some visually obvious artifacts, including a bright foreground star in the center of the field, as well as some claws and wisps due to scattered light from nearby bright objects.  

As with NIRCam, the culling criteria identified $\sim98$\% of the objects as non-stellar artifacts.  The resulting stellar field is sparsely populated, which is expected due to its location at $\sim$ 5 half-light radii.  

Figure~\ref{fig:m92_cmd_niriss} shows the NIRISS CMDs of M92 for all objects (left) and stars (right).  The majority of non-stellar objects rejected by the catalog culling are faint sources near the bottom of the CMD.  The resulting stellar CMD in the left panel shows a clear lower MS of M92.  The NIRISS CMD spans a smaller dynamic range in luminosity compared to NIRCam.  At the bright end, the CMD reaches the top of the MS kink, not the oldest MSTO.  This is likely due to saturation effects owing to different detector characteristics. The NIRISS stellar CMD has stars nearly as faint as those in NIRCam, but the scatter is visually much larger at the bottom of the MS.  This may be due to the lower throughput of NIRISS.  Additionally, as discussed in \S \ref{sec:dolphot_photometry}, the NIRISS WebbPSF models are marginally worse matches to the observed PSFs (i.e., the models have too much centrally concentrated light), which results in larger PSF adjustments from DOLPHOT and a larger systematic uncertainty floor of $\sim0.01$~mag per filter, which is an order-of-magnitude larger than the PSF-based uncertainties for NIRCam.  The net result is that the increased width of the MS in M92 is at least in part driven by the model PSFs.

\begin{figure*}[th!]

\epsscale{1.2}
\plotone{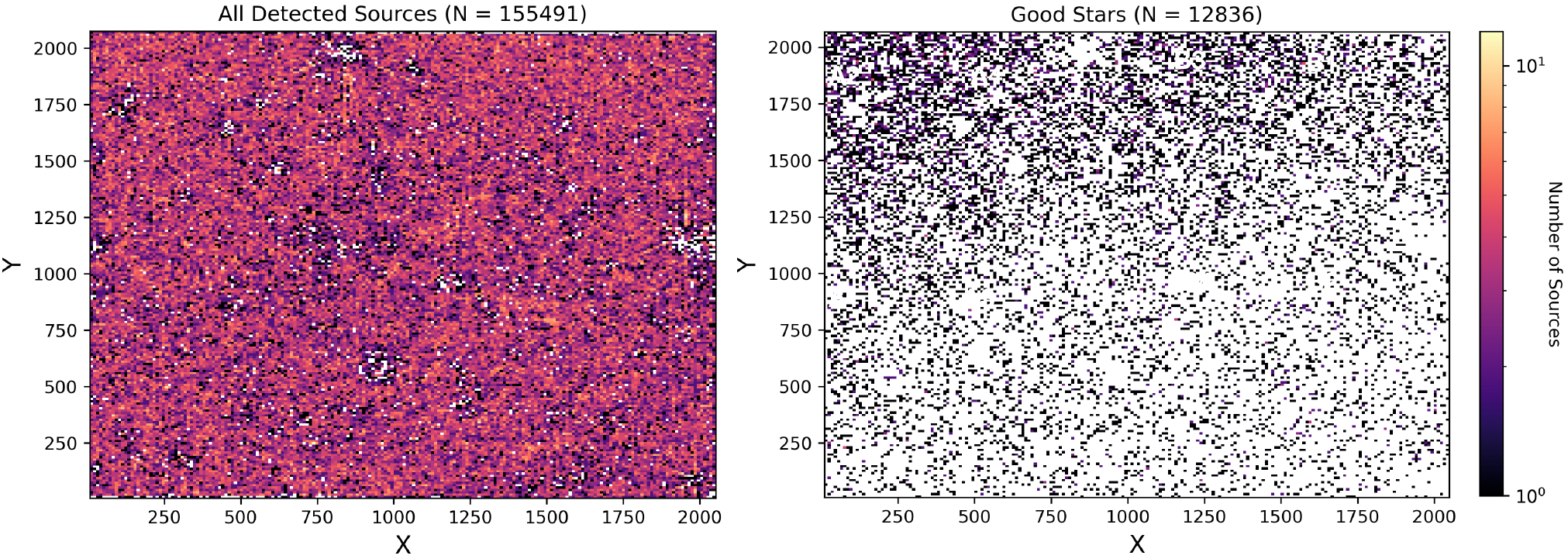}

\caption{The spatial distribution of objects and stars detected by DOLPHOT in our NIRISS imaging of WLM.   \textbf{Left:} A density map ($\sim50\times50$ pixel bins) of the $\sim 1.6\times10^5$ objects reported in the raw DOLPHOT catalog.  The spatial distribution of detected objects is fairly uniform.  `Holes' in the spatial distribution mainly correspond to saturated pixels that were entirely masked by DOLPHOT.   \textbf{Right:} The density map of the $\sim1.3\times10^4$ stars that passed the catalog culling criteria listed in \S~\ref{sec:cull}. There is a modest spatial gradient that increases in the direction of the main body of the galaxy.}
\label{fig:wlm_spatial_niriss}
\end{figure*}

Table~\ref{tab:ers_asts} includes summary statistics of the NIRISS M92 SW ASTs.  The recovered ASTs show no bias and minimal scatter, much like the NIRCam ASTs.  The 50\% completeness limits of the NIRISS field are $\sim2$ magnitudes fainter ($m_{F090W}=28.1$, $m_{F150W}=27.4$) than the NIRCam field, which is the result of no crowding in the NIRISS field, compared to modest crowding in the center of the NIRCam field.  The density of objects and stars in the NIRISS field is $\sim4-10$ times less than it is in the NIRCam field (Table~\ref{tab:ers_density}).

\subsection{WLM}

\subsubsection{NIRCam}

Figure~\ref{fig:wlm_spatial} shows the spatial distribution of objects (left) and stars (right) from our DOLPHOT photometric reduction of WLM.  In total, DOLPHOT finds $\sim1.7\times10^6$~objects in the WLM field, yielding a typical density of $\sim49$~objects per sq. arcsec.  Of these, $\sim4.6\times10^5$ ($\sim27$\%) pass the culling criteria, yielding a stellar density of $\sim13$ stars per sq.\ arcsec, which is the highest density of our ERS fields.  

The WLM observations are oriented such that chips A1 and A2 are closest to the center of the galaxy, while B1 and B2 are farthest away.  This orientation produces a spatial gradient from A1 (highest density) to B1 (lowest density) that is clearly visible in both the object and stellar spatial maps.  As with M92, the object map shows several artifacts (foreground stars, background galaxies) that are largely rejected by the culling criteria.  The inter-module and inter-chip gaps are not populated due to our choice in dithers and culling criteria.

Figure~\ref{fig:wlm_nircam_cmd} shows select NIRCam SW and LW WLM CMDs.  Panel (a) plots the F090W$-$F150W CMDs for all objects (left) and stars (right). The majority of sources removed by the catalog culling are faint, low S/N objects.  The resulting stellar CMD of WLM is the deepest ever obtained for a galaxy outside the virial radius of the MW.  It's remarkable for its depth and precision.    Many of the features in the SW CMD are similar to a previous analysis of WLM with HST/ACS \citep{albers2019} and are discussed in more depth in our team's SFH paper of WLM \citep{mcquinn2023}.  Here, we briefly summarize CMD features.   At the bright end, we see a clear young MS population indicting the presence of recent star formation.  Slightly redder than the MS, is the  blue core helium burning sequence (BHeB).  Part of this sequence falls in the instability strip and these stars appear as Cepheids, some of which have been targeted from the ground over the last few decades \citep[e.g.,][]{sandage1985, pietrzynski2007}. For redder bright stars, we see a well-defined population of asymptotic giant branch (AGB) stars that are located above a clearly defined tip of the red giant branch (TRGB).  This AGB star population is analyzed in detailed in \citet{boyer2024}.  The RGB is narrow and well-populated.  A mixture of AGB stars and red core helium burning stars are located at slightly bluer colors than the RGB.   There is a prominent, tight red clump (RC) at $m_{F090W} \sim 25$ along with a clear horizontal branch.  Vertically extending from the red clump is the red helium burning sequence, the brightest of which are considered red supergiants.  The CMD extends $\sim2$~magnitudes below the oldest main sequence turnoff.  

Panel (b) shows the LW CMDs of the WLM NIRCam field.  The culling criteria have left a fair number of low S/N objects on the stellar CMD, indicating that improvements could be made by including the LW filters in the culling.  Readers who wish to explore this are encouraged to download our photometry from MAST.

As expected, the RGB in the stellar CMD is narrow in this filter combination as they are both well into the Rayleigh-Jeans tail of RGB stellar flux distributions.  There is a prominent, bright AGB star population. The CMD begins to broaden substantially below the red clump.  Though the LW CMD doesn't reach the oldest MSTO, it is nevertheless remarkable as it is the deepest medium band CMD of a galaxy outside the MW satellites.

Panel (c) shows an illustrative SW$-$LW CMD.  Here, the culling criteria do an acceptable job of removing non-stellar objects, though including F250M-specific criteria should improve the faint end source classification.  The CMD features are generally similar to those shown in panel (a) though the CMD is not as tight in general or quite as deep.  Culling criteria tailored to F250M, along with ASTs, are necessary to determine if the oldest MSTO is brighter than the 50\% completeness limits.

The SW AST results for WLM are summarized in Table~\ref{tab:ers_asts}.  The 50\% completeness limits of F090W (28.7) and F150W (27.7) are just below the oldest MSTO.  Because the culling criteria were designed for purity, and not completeness, relaxing the sharpness will extend the completeness limits fainter.  Such decisions need to be driven by the science. For example, star formation history measurements may be able to tolerate a decrease in purity in exchange for fainter completeness limits \citep[e.g.,][]{mcquinn2023}.

\begin{figure}
    \centering
\plotone{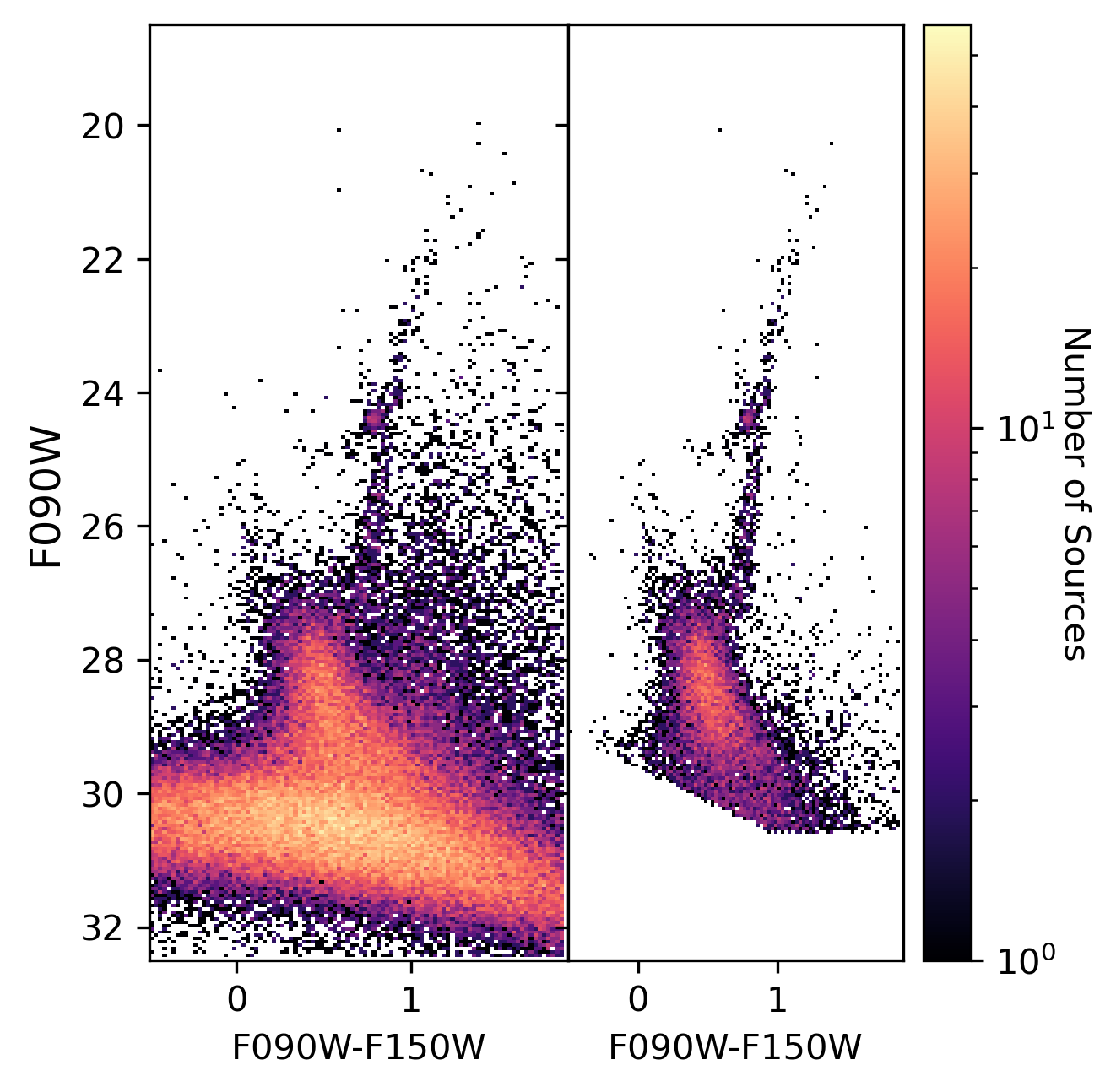}
    \caption{The NIRISS CMD of WLM.  The left panel shows all objects detected, the right panel shows the stars that passed the culling criteria. The NIRISS CMD has a similar depth to the NIRCam CMD, albeit with increased scatter. In part, the scatter is due to the lower sensitivity of NIRISS in these filters, along with the less accurate NIRISS WebbPSF models.  As expected, due to its location outside the main body of the galaxy, the NIRISS CMD lacks a  stellar population younger than a few Gyr and is much more sparsely populated.}
    \label{fig:wlm_cmd_niriss}
\end{figure}

\subsubsection{NIRISS}

Figure~\ref{fig:wlm_spatial_niriss} shows the spatial distribution of objects (left) and stars (right) in the NIRISS field of WLM.  The objects are generally distributed uniformly, with a handful of low density regions owing mainly to saturated pixels.  The culling criteria removes $\sim90$\% of the objects from the field, leaving a sparse stellar distribution, which is consistent with this field's location in WLM's stellar halo.  There is a slight gradient in the field toward the upper left portion of the field, which is also the direction of the center of the galaxy.    The object and stellar densities are 8.9 and 0.7 per sq.\ arcsec, indicating that this is not a crowded field.

Figure~\ref{fig:wlm_cmd_niriss} shows the CMD of objects (left) and stars (right) for the WLM NIRISS field.  The culling criteria remove much of the contamination around the RGB and below the oldest MSTO, producing the deep and clean stellar CMD.  The stellar CMD lacks stars younger than at least 1-3 Gyr, due to its location in the stellar halo.  Otherwise, it exhibits many of the expected features of old and intermediate age populations (e.g., RGB, RC).  The CMD extends well-below the oldest MSTO, similar to the NIRCam CMD.  The RGB, RC, and MSTO appear slightly broader on the NIRISS CMD compared to NIRCam.  This is unlikely to be due to a more complex stellar population, and instead may reflect differences in NIRISS and NIRCAM, such as lower throughput and slightly less accurate PSF models for NIRISS.

Table~\ref{tab:ers_asts} summarizes the AST results for the WLM NIRISS field. As with the other field, the ASTs show no bias and little scatter, indicating that they are well-recovered.  The 50\% completeness limits are $m_{F090W}=29.3$ and $m_{F150W}=28.5$, which are 0.6 and 0.8~mag deeper than the same filters in NIRCam.  This is because the NIRISS field is located in the much less crowded stellar halo.

\begin{figure*}[ht!]

\epsscale{1.2}
\plotone{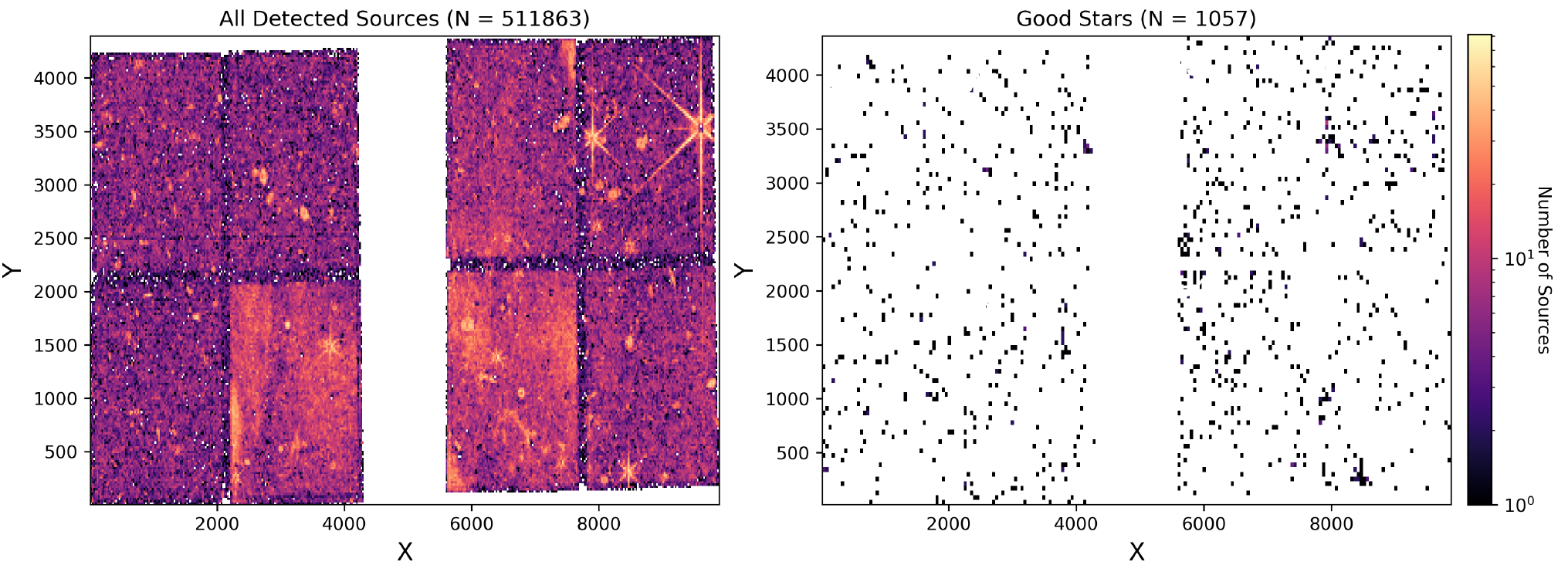}

\caption{The spatial distribution of objects and stars detected by DOLPHOT in our NIRCam imaging of \dracoii.   \textbf{Left:} A density map ($\sim50\times50$ pixel bins) of the $\sim 5.1\times10^5$ objects reported in the raw DOLPHOT catalog.  The spatial distribution of objects shows a number of features, including saturated foreground stars and background galaxies.  The other prominent features are the diffuse object overdensities that are present in chips A3, B3, and B4.  These are the result of persistence from the previous program (1022), which was staring at/near Mars, Jupiter, and Saturn for $\sim$12 hours in order to test the fine guidance sensor. \textbf{Right:} The density map of the $\sim1.1\times10^3$ stars that passed the catalog culling criteria listed in \S \ref{sec:cull}.  These criteria removed the vast majority of obvious artifacts (e.g., corresponding to saturated stars, diffraction spikes) as well as eliminated most of the contamination from persistence.}
\label{fig:draco_spatial}
\end{figure*}

\subsection{Draco II}
\label{sec:draco}

Figure~\ref{fig:draco_spatial} shows the spatial distribution of objects (left) and stars (right) for our NIRCam observations of \dracoii.  The object density plot shows many familiar artifacts including bright foreground stars, chip gaps, and background galaxies.  Additionally, there are large, diffuse overdensities that cover chips A3, B3, and B4.  These features are also present in the images themselves and are the result of persistence.  Prior to our observations, program 1022 spent $\sim12$ hours testing the fine guidance sensor's ability to track moving objects near Mars, Jupiter, and Saturn.  This resulted in back-to-back observations of some of the brightest objects in the Universe, followed by one of the faintest objects in the Universe.

As shown in the right panel, the culling criteria does an excellent job of removing these overdensities, along with the other artifacts. The result is an extremely low density stellar field with just 0.03 stars per sq.\ arcsec, which is typical of an ultra-faint dwarf galaxy.  \citet{bagley2023} report persistence in the CEERS data from the same Solar System program and develop a routine to mask out pixels affected by persistence.  In our case, this would result in $\sim$1/3 of our field not being analyzed at all.  While this provides a suitable solution, another may be to specify that observations should not be scheduled when extreme persistence may be a problem.

\begin{figure*}[ht!]
\gridline{\fig{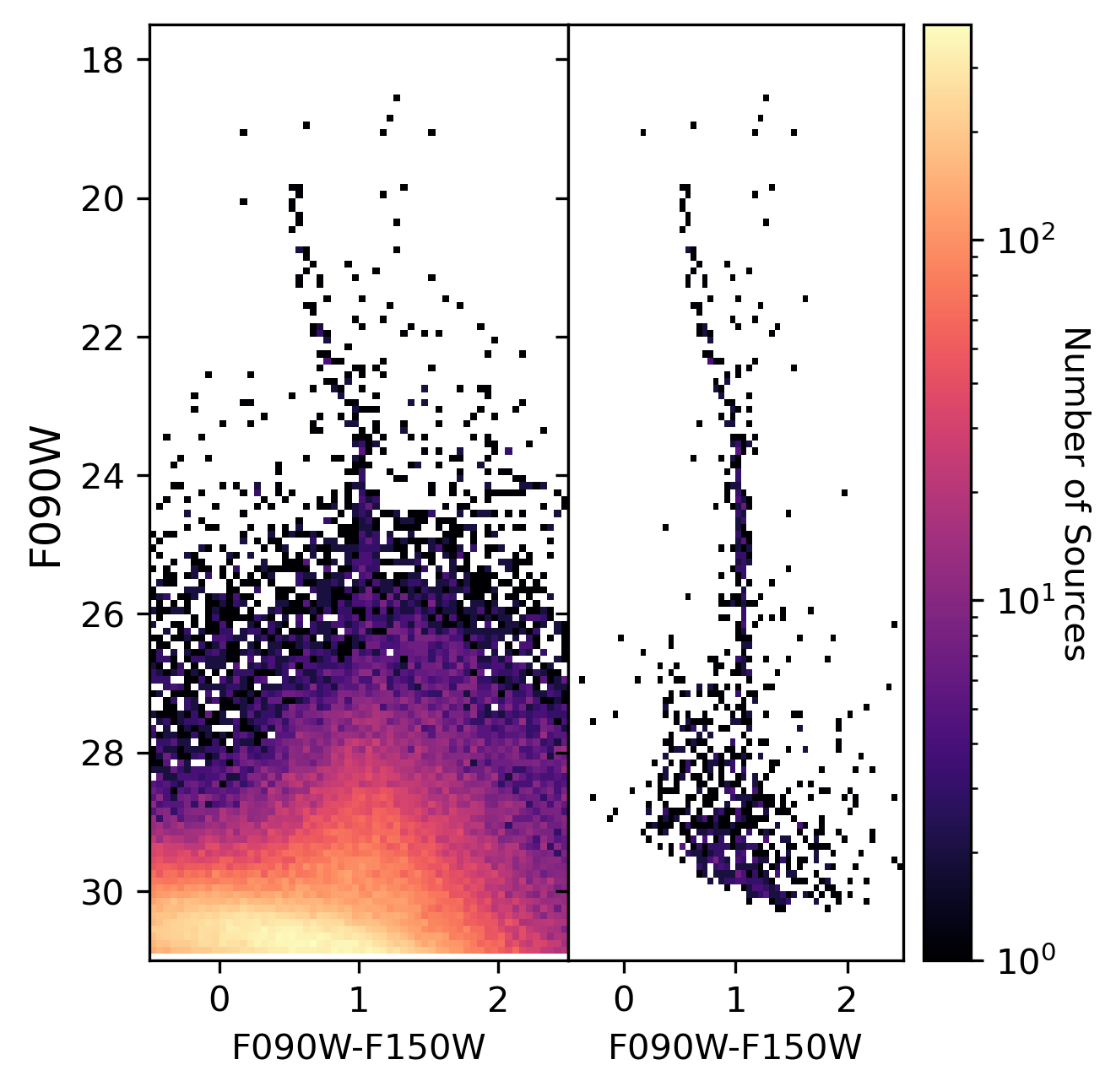}{0.34\textwidth}{(a)}
\fig{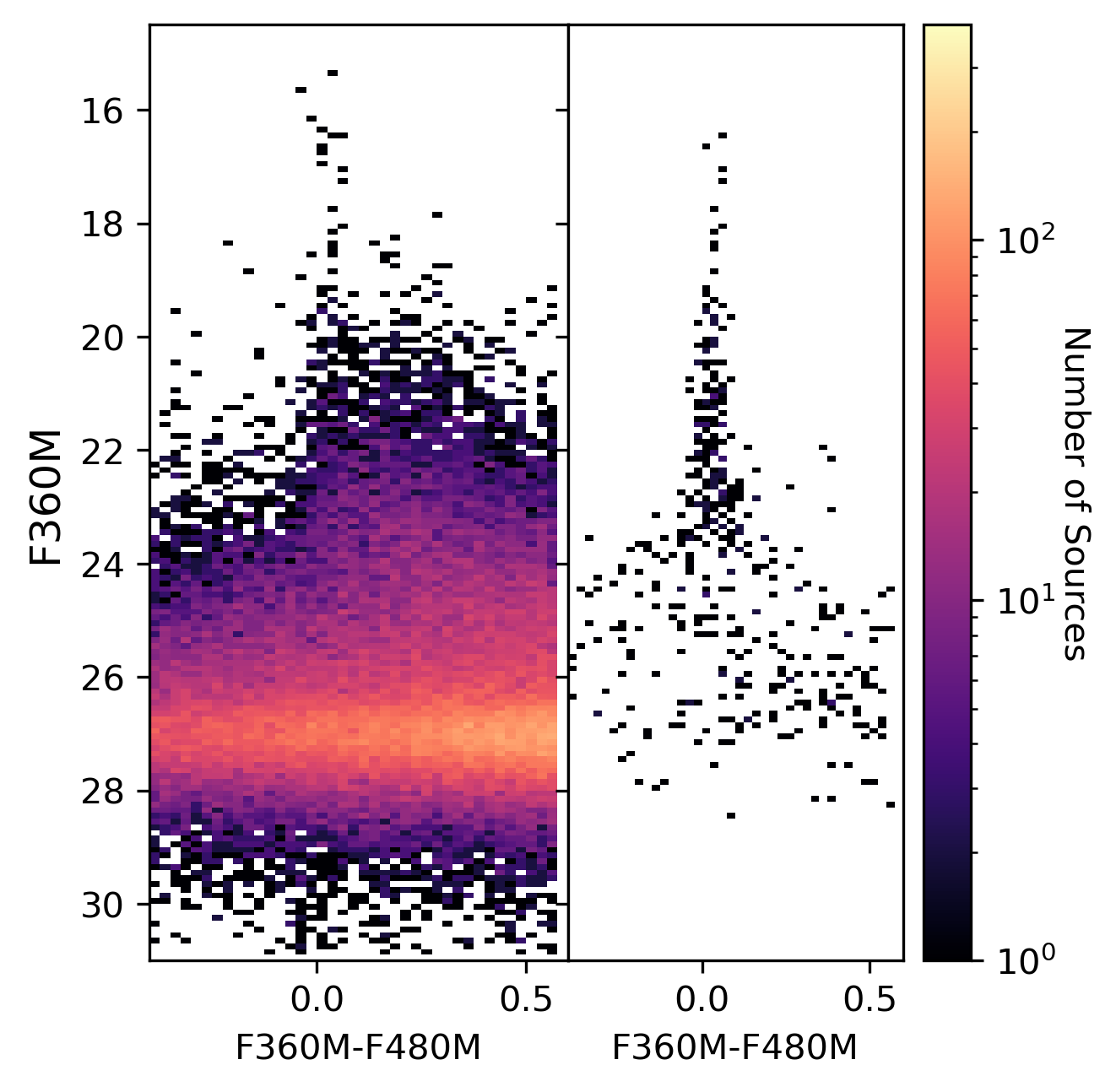}{0.34\textwidth}{(b)}
\fig{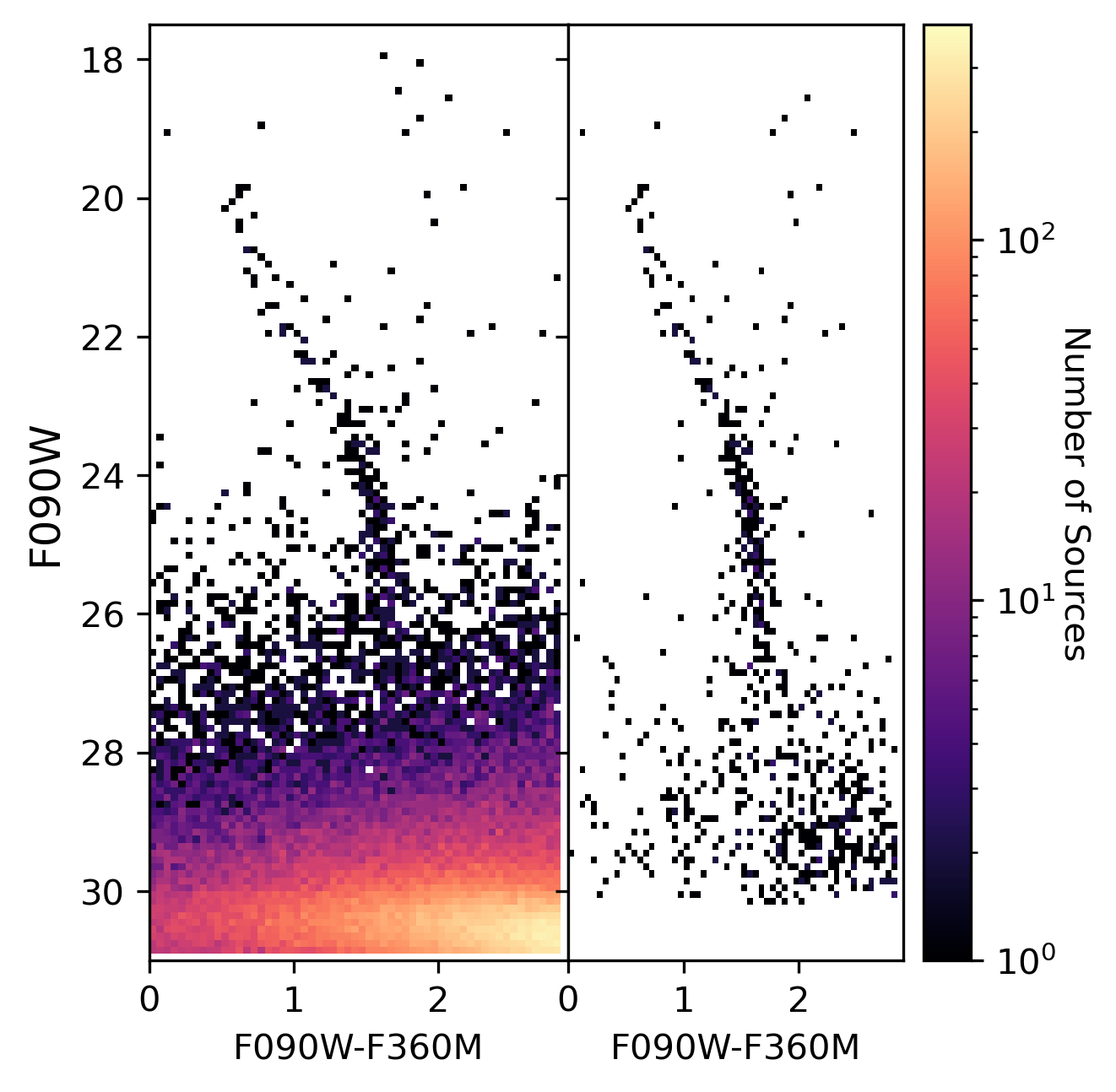}{0.34\textwidth}{(c)}}
\caption{NIRCam CMDs of \dracoii\ in select filter combinations.    Panels (a), (b), and (c) show the CMDs for the SW (F090W$-$F150W), LW (F360M$-$F480M), and an example SW/LW (F090W$-$F360M) filter combination.   The SW CMD of \dracoii\ (panel a) is the deepest ever (i.e., it reaches the lowest stellar mass) constructed for a galaxy outside the MW itself.  It includes a clearly defined MS Kink.   Panel (b) shows the LW-only CMD, which is essentially vertical at a color of $\sim0$ due to the lack of temperature sensitivity in the F360M$-$F480M filter combination. Panel (c) shows an example SW-LW CMD (F090W$-$F360M), which extends below the MS kink before photometric scatter washes out the stellar sequence.  This depth is impressive for a medium-band LW filter.  For both CMDs that include the LW filter, specific LW culling criteria may reduce some of the noise in the CMD, particularly at the faint end.  
\label{fig:draco_nircam_cmd}}
\end{figure*}

Figure~\ref{fig:draco_nircam_cmd} shows select CMDs of \dracoii.  Panel (a) shows the SW CMD in which application of the culling criteria produces a deep, fairly clean CMD of \dracoii.  The CMD extends from the oldest MSTO to beyond the bottom of the stellar sequence.  The MS kink is clearly visible at F090W$\sim23$.  The large scatter at the bottom of the CMD is the result of confusion between background galaxies, stars, and the gap between stars and brown dwarfs. Even with cuts designed for purity, we are not able to readily discern between stars and compact galaxies at such faint magnitudes. This is the deepest CMD (i.e., it reaches the lowest mass main sequence star) ever constructed of a galaxy outside the MW.

Panel (b) shows the LW of \dracoii.  The SW culling criteria drastically reduce the number of contaminants, leaving a clear MS in the right hand panel.  Further contamination, particularly near the faint end, could possibly be removed by adding LW-specific culling criteria. The F360M-F480M color provides little leverage on temperature, resulting in a nearly vertical MS.

Panel (c) shows an example SW and LW CMD (F090W-F360M). The SW culling criteria do a reasonable job removing contamination down to F090W$\sim26$, below which there is a noticeable increase in scatter.  This scatter is likely due to the low SNR of the F360M data at such faint magnitudes as well as the lack of an LW-specific culling criteria.  The stellar CMD shows a clear lower MS, including the MS kink.  The F360M filter was taken specifically for its metallicity sensitivity and analysis of this CMD could, in principle, provide one of the tightest constraints on whether \dracoii is a \textit{bona fide} UFD or GC, which remains an open question in the literature \citep[e.g.,][]{baumgardt2022, fu2023}, by determining if its metallicity distribution function has a statistically significant spread.

Table~\ref{tab:ers_asts} lists the AST properties for this NIRCam field.  The 50\% completeness limits are $m_{F090W} = 29.6$ and $m_{F150W} = 28.3$, which are nearly at the bottom of the CMD.  The NIRISS field did not have enough bright stars to align properly in DOLPHOT and we therefore do not discuss it.

We note that although the culling criteria appear to do a good job of removing the contamination due to persistence, it is unclear how many stars in \dracoii\ were also removed.  In general, it may be advisable in proposal planning to request that observations of resolved galaxies be scheduled such that persistence is unlikely to be an issue.  If this had been WLM instead of \dracoii, it is possible that a significant number of stars may have been lost in the persistence-induced noise.

\begin{figure*}[ht!]
\plotone{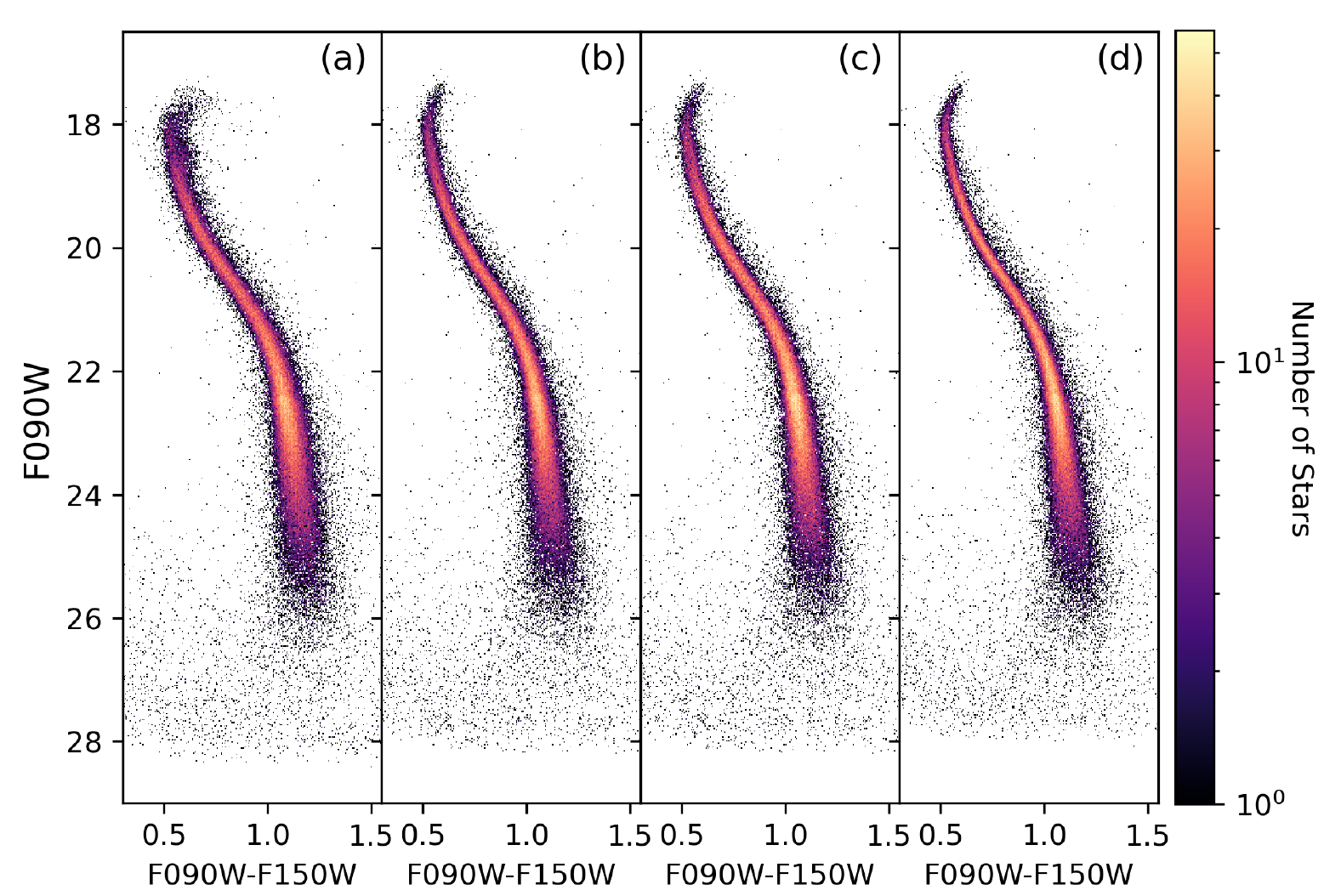}
\caption{An illustration of improvements in DOLPHOT NIRCam photometry since the publication of our survey paper in early 2023 demonstrated by showing F090W-F150W CMDs of M92 from various DOLPHOT runs in the past year.  Panel (a) shows the CMD from our ERS survey paper \citet{weisz2023a} from January 2023.  It was constructed using all exposures of M92, including the anomalous 3rd exposure, and used \jwst\ calibrations and WebbPSF models from late 2022.  Panel (b) shows photometry from the same time frame, only without the 3rd exposure.  Panel (c) shows the CMD of M92 (no 3rd exposure) using the older WebbPSF models (v1.1) and calibration data, but with updated NIRCam zero points released in Fall 2023.  Panel (d) shows the M92 CMD published in this paper (with no 3rd exposure), which includes updated WebbPSF models and very recent calibrations (e.g., flat fields) and zeropoints.  From left to right, the CMDs have noticeably less scatter, tighter stellar CMD sequences (e.g., MSTO, MS kink), more stars, and improved depth.  \label{fig:m92_improvements}}
\end{figure*}

\section{Discussion}
\label{sec:discussion}

\subsection{Evolution of \jwst\ DOLPHOT Photometry}
\label{sec:nircam_improve}

Our knowledge of \jwst\ and its instruments has greatly improved since our team's ERS data was acquired in mid-2022.  Among the improvements during this time are more accurate calibrations (e.g., flat field, zero points), better data quality masking, and more realistic model PSFs.  During the course of our ERS program, we have continued to incorporate these changes into DOLPHOT.

Figure \ref{fig:m92_improvements} illustrates the impact of these revisions on the SW CMD of M92.  Panel (a) shows the SW CMD of M92 that was originally published in our ERS survey paper \citep{weisz2023a}, which includes all 4 exposures.  Panel (b) was reduced with the same DOLPHOT configuration as panel (a), but without the anomalous 3rd exposure.  CMDs in both panels (a) and (b) were constructed using images produced by the the \jwst\ pipeline version with \texttt{CAL\_VER}$=$1.9.3,\texttt{CRDS\_VER}$=$11.16.18, and \texttt{CRDS\_CTX}$=$jwst\_p1063.pmap, as well as PSF models using WebbPSF version 1.1.1.  

The CMD in Panel (c) used the same setup as above, but with different zeropoints.  Specifically, they are from \texttt{CRDS\_CTX}$=$jwst\_p1126.pmap, whichas released in Fall 2023.  These zero points were applied to the photometry after it was already run. A notable improvement in panel (c) was a reduction in chip-to-chip photometric offsets, which we observed to range from 0.02 to 0.1 in all filters in all previous version of our photometry (i.e., panel b).  

Finally, panel (d) shows the CMD presented in this paper, with the most up-to-date PSFs and calibrations.  Some of the updates include flat fields, zero points, the switch from Vega to Sirius as a reference star, and WebbPSF models that model interpixel capacitance and charge diffusion.  WebbPSF v1.2 was released in late 2023.

Visually, there is clear and dramatic improvement in these CMDs over time. The progression from panel (a) to (d) is one of tighter sequences, less scatter, improved definition of the MSTO, and greater depth.  Our most recent CMD also contains more stars than previous versions.  

\begin{figure}[th!]
\plotone{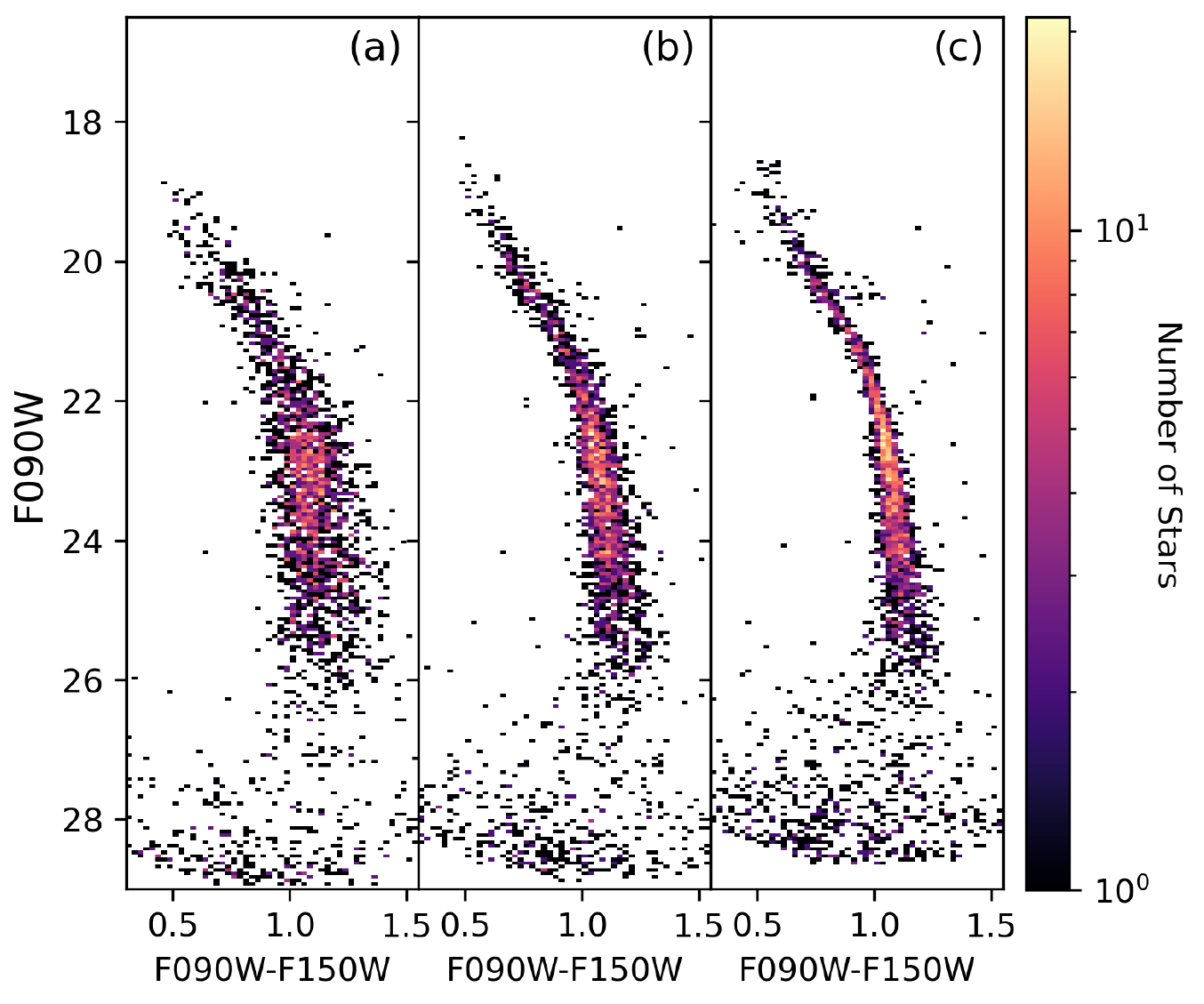}
\caption{Improvements in DOLPHOT NIRISS photometry, illustrated by CMDs of M92.   Panels (a) and (b) are the NIRISS CMDs released as part of the ERS survey paper in January 2023 \citep{weisz2023a}; with the CMD constructed from 4 exposures shown in panel (a) and the CMD constructed without the 3rd exposure in panel (b).  Panel (c) shows the NIRISS CMD from this paper, which includes update WebbPSF models and \jwst\ NIRISS calibrations.  It does not include the 3rd exposure.  The CMD in panel (c) has less scatter, tigther sequences, more stars, and greater depth than our previous NIRISS CMDs. \label{fig:m92_improvements_niriss}}
\end{figure}

Figure \ref{fig:m92_improvements_niriss} shows an even more dramatic improvement over the same time and parameter range.  Our initial NIRISS CMD (panel a) exhibited a tremendous about of scatter due in large part to the effect of the 3rd exposure.  The removal of the 3rd exposure (panel b), significantly reduced the scatter in the CMD.  Panel (c) shows the current NIRISS CMD of M92, with no 3rd exposure.  Relative to panel (b) it has a tighter main sequences and contains more stars.  These improvements were almost entirely the result of improved WebbPSF models.  Previous WebbPSF models had too much light concentrated in the central pixel compared to observations.  The current NIRISS WebbPSF models are still slightly too sharp, but this only affects the photometry at the level of $\sim0.01$~mag, whereas the previous generation introduced a scatter of $\sim0.09$~mag.

\subsection{Point Spread Function Time Variability}  
\label{sec:time_variations}
Space-based telescopes are typically characterized by a much higher degree of PSF stability than ground-based facilities.  However, even space telescopes exhibit some PSF time dependence, due to, for example, thermal variations or small impacts.  \hst\ is known to exhibit such effects \citep[e.g., optical telescope assembly breathing;][]{hasan94} and they are generally small and stable enough to be corrected for by the PSF model adjustments performed by DOLPHOT.  \jwst\ is expected to have similar temporal changes in the PSF (e.g., see Sec 6.2 of \citealt{McElwain2023}).  Here, we undertake a preliminary characterization of the effects of temporal variations in the PSF on the DOLPHOT photometry of the ERS targets.

\begin{figure*}[ht!]
\epsscale{2}

\plottwo{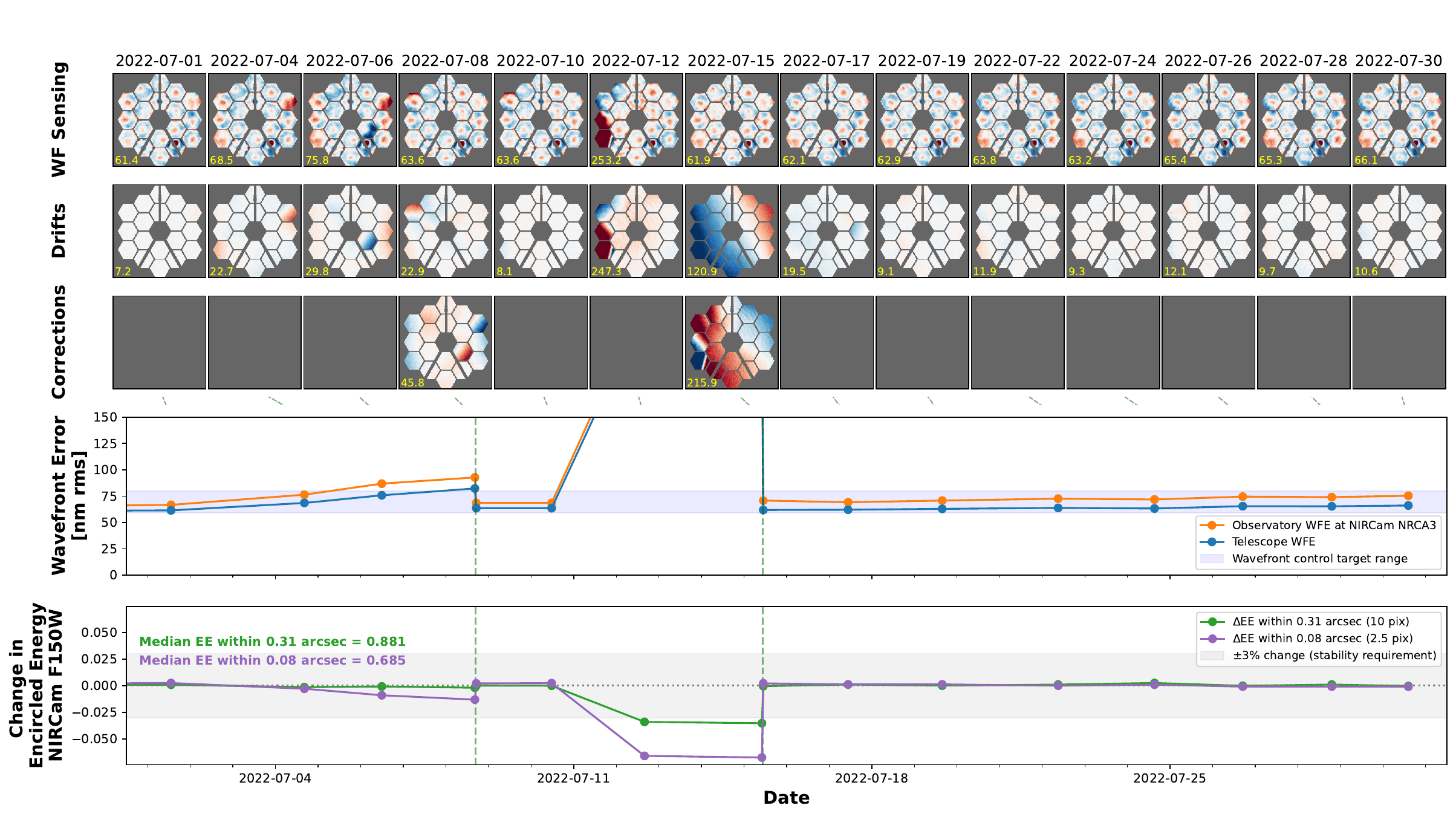}{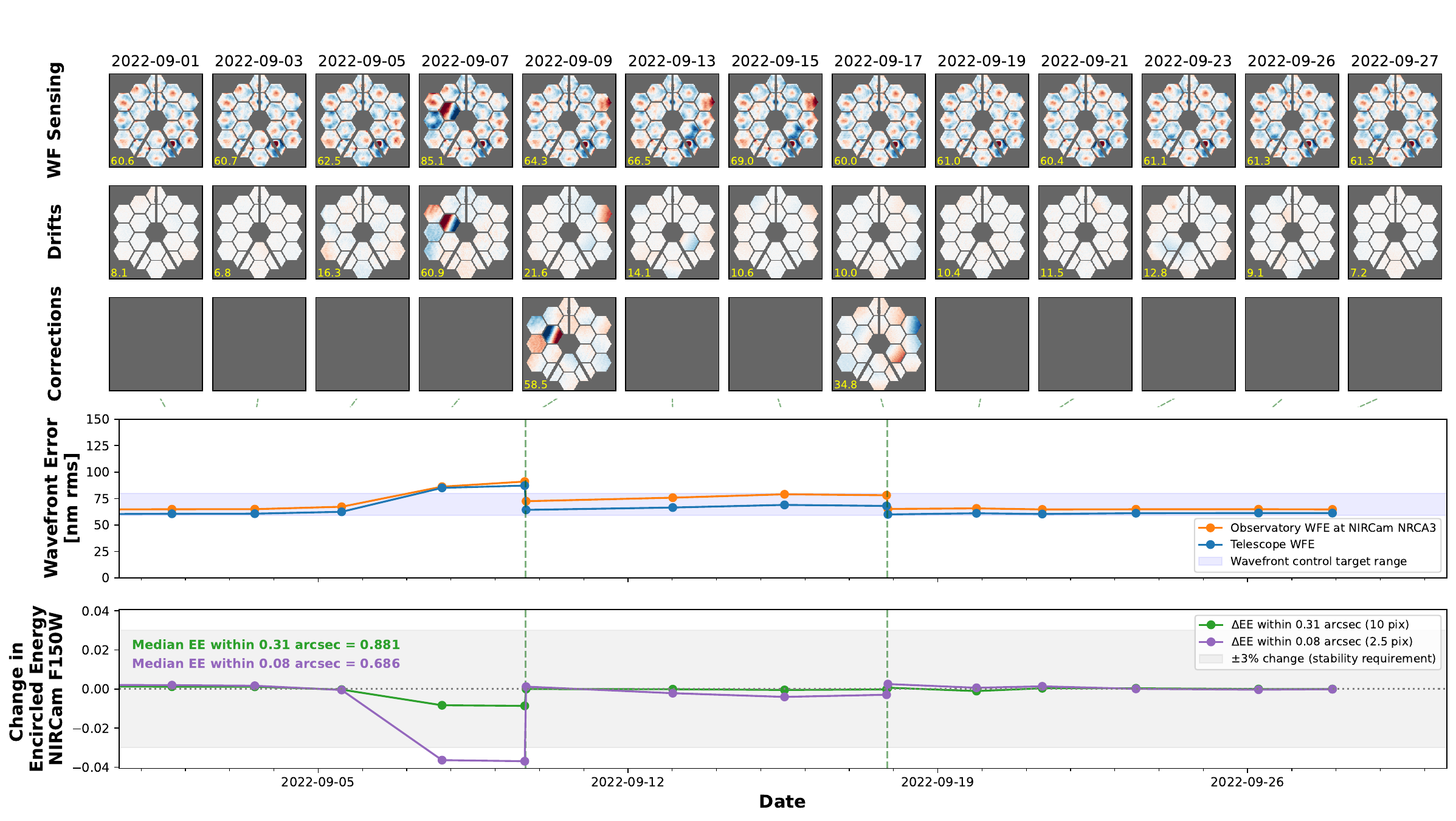}
\caption{Figures output by WebbPSF that show variations in the telescope alignment, wavefront, and encircled energy as a function of time over the period.  At select epochs, we compute the effects of small, large, and no perturbations to the telescope stability on DOLPHOT photometry.}
\label{fig:Alignment}
\end{figure*}

Figure~\ref{fig:Alignment} shows the time-series wavefront sensing measurements for \jwst's optical telescope element (OTE), along with related encircled-energy variations in the NIRCam F150W PSF, for the months of July (top plot) and September (bottom plot) 2022 as generated by WebbPSF.   For most of the measured epochs, \jwst's optical performance shows remarkable stability, with minimal deviations from commissioning alignment. However, sporadic events can occur when the telescope drifts away from nominal performance. While corrections to the mirror segment positioning are rapidly issued to bring the telescope back to commissioning alignment, observations taken before the corrections are applied will likely present significant variations from nominal PSF models.

Within the twelve months period spanning June 1st, 2022 to May 31st, 2023, 14 such events occurred. Deviations from nominal performance lasted between two and seven days before corrections were issued. The largest event recorded so far occurred between July 11th 2022 and July 15th 2022 (top plot of Fig.~\ref{fig:Alignment}), with changes to the encircled-energy of various filters changing more than 5\% at a 10 pixel radius. 

However, the majority of the alignment anomalies were much smaller. The bottom panel of Fig.~\ref{fig:Alignment} shows a typical example of such an event, which, in this case, occurred between September 6th 2022 and September 10th 2022.  Resultant changes to the PSF were within mission stability requirements.

As the majority of these events are related to thermal settling of the spacecraft \citep[e.g.,][]{McElwain2023}, they primarily occurred in the first six months of scientific operations. In fact, since November 2022, only two such misalignments have occurred, both of them with minimal deviations from nominal performance. The outlook for \jwst's optical stability is therefore very promising. 

Nevertheless, is important to quantify the impact of time-dependent PSFs on the photometry, especially for datasets acquired early in Cycle 1. To do so, we computed two alternative PSF grids for NIRCam, using OPD maps corresponding to the July ``large event'' (\texttt{R2022071502-NRCA3\_FP1-1.fits}; July 15th, 2022) and to the September ``small event'' (\texttt{R2022090902-NRCA3\_FP1-1.fits}; September 9th, 2022). We also calculated a third grid which corresponds to nominal alignment changes in the telescope two months after our official PSFs OPD (``O2022092302-NRCA3\_FP1-1.fits''; September 23rd, 2022). We term this last test the ``no event'' case.  The no event case is meant to test how normal operational variations in the telescope (e.g., mirror alignments, thermal effects) manifest in DOLPHOT PSF photometry under the assumption that the PSF is computed at one epoch but applied to data taken at an epoch 2 months later.  We then re-ran DOLPHOT on our three ERS targets, using these alternative grids, and compared the differences in the photometry.  

The DOLPHOT-generated catalogs from the two epochs are spatially cross-matched by (a) only considering stars with SNR$>50$ in both epochs and (b) requiring their spatial coordinates to match within $0.15$~pix.  For the large event, we were only able to adequately match sources with a much larger radius of $2$~pix. We discuss the implications of the spatial matching radius below.  We use these high SNR stars cross-matched between the two epochs to assess differences in the photometry.  As a point of reference, we also compare them to the expected scatter from ASTs of images analyzed with PSFs at the same epoch, i.e., our nominal photometry.

\begin{figure}[t!]

\plotone{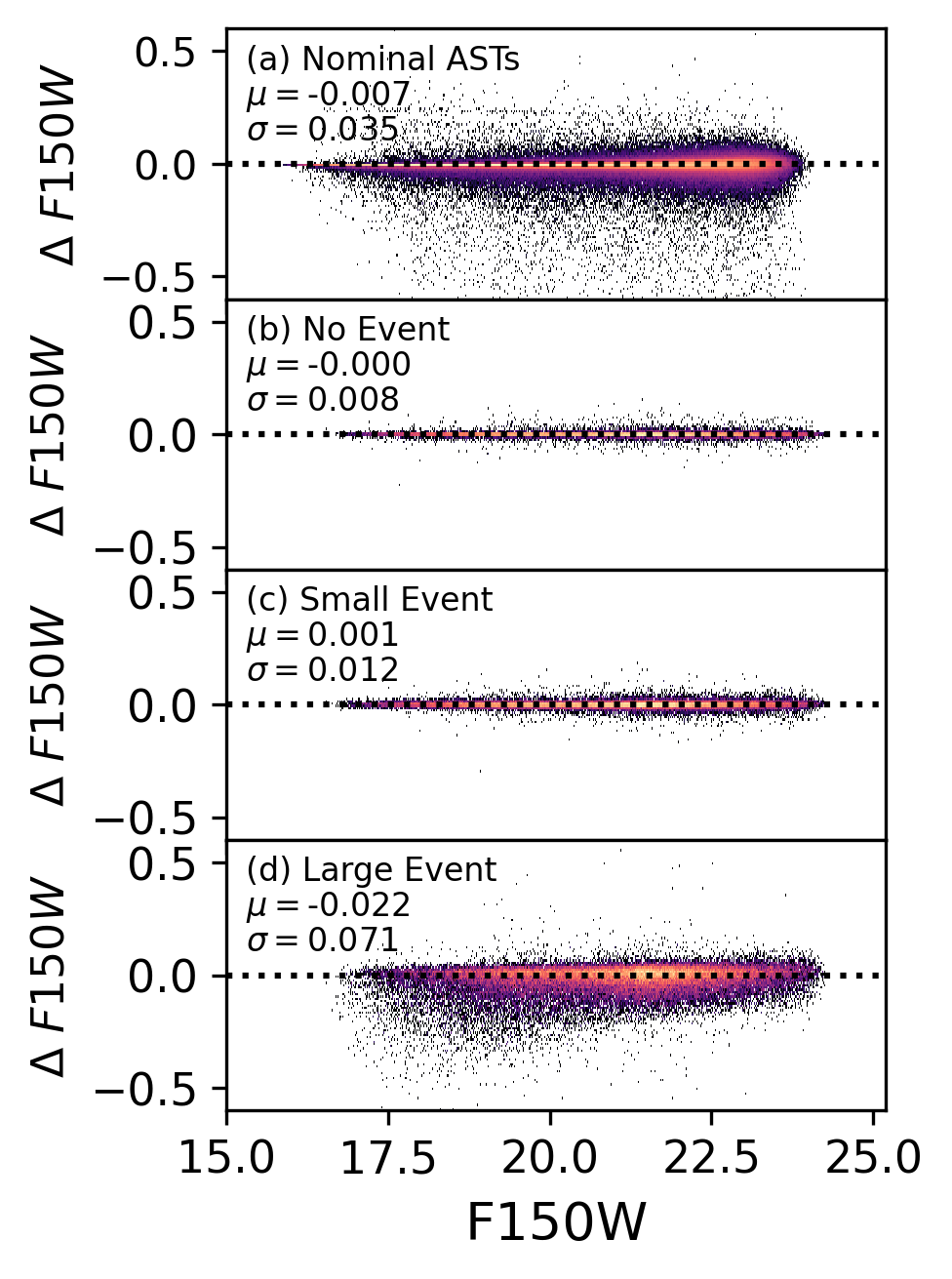}
\caption{An illustration of the effects of temporal variations on the photometry of M92.  We show scatter from the SNR$>50$ M92 ASTs in F150W in panel (a) as a point of comparison.  The other panels show the no event case (panel b; i.e., normal temporal variations), a small misalignment event (panel c), and a large misalignment event (panel d).  Each of these panels shows the difference in F150W magnitudes of the same high SNR stars (SNR$>50$) for our fiducial photometry and photometry computed using WebbPSF models from an epoch corresponding to an event.  For normal operation (panel b) and small misalignment events (panel c) there is very little differnce in the photometry. For the a large misalignment event, a non-negligible amount of scatter can be present and should be included in the error budget.   Large events may also impact the astrometry.  See \S~\ref{sec:time_variations}
 and Table~\ref{tab:time_variations} for more details. Fortunately, such large events appear to be rare.
\label{fig:Time_WLM}}
\end{figure}

For each test, we compare properties of the cross-matched DOLPHOT photometry to the nominal photometry, i.e., the photometry presented in \S \ref{sec:ers_photometry}.  For the nominal case, we use the ASTs with SNR$>$50 to assess the bias and scatter, which serve as a reference point by which to assess the effects of time variations on the photometry. To illustrate expectations from the ASTs, we show the expected bias and scatter for M92 in the F150W filter in the top panel of Figure~\ref{fig:Time_WLM}.  We see the bias is smaller than the scatter and is consistent with zero.  The amplitude of the scatter increases as expected for ASTs.

The next 3 panels in Figure~\ref{fig:Time_WLM} show an example of how the M92 NIRCam F150W photometry compares between our nominal catalog and the three types of events we consider. Specifically, we plot the difference in F150W magnitudes between the two sets and compute the mean and scatter for all stars with SNR$>50$.  

Because we are comparing photometry of identical stars between the epochs, the expectation is that the difference in magnitude should always be zero.  The only variable in the reduction is the PSF library, thus any differences we find are solely due to variations in the PSF.  

For the no event case, we find no bias ($\mu=-0.001$) and a very small scatter ($\sigma=0.008$). This indicates that while the photometry is not identical between the epochs, the differences are small and on order of systematics introduced by the PSF models at the same epoch.  These effects are also smaller than the noise reported by the ASTs, shown in panel (a).

The small event case (panel c) tells a similar story.  The bias and scatter are small, though the scatter is not entirely negligible.  Specifically, in this case, an accurate characterization of the noise would require adding $\sigma=0.012$~mag in quadrature to other sources of noise.

The bottom panel of Figure~\ref{fig:Time_WLM} shows the results for a large event.  In this case, the mean difference in the photometry is small, but non-zero ($\mu = -0.02$~mag), and the scatter $\sigma = 0.07$~mag) is larger than in the no event and small event cases. The scatter is a factor of $\sim2$ larger than the noise reported by the ASTs and a factor of $\sim3.5$ larger than the noise from the Poisson noise in the photometry (i.e., SNR$>50$ translates to a photometric error of $<0.02$~mag).  In the large event case, the time variations in the PSF are actually the dominant source of photometric uncertainty and would need to be included in any subsequent modeling of the data.

Beyond the addition of significant photometric noise, we also found that the large event made the astrometry less robust.  Specifically, in order to match stars between the large event and nominal catalogs, we had to expand the pixel matching radius from 0.15 to 2~pix, over an order-of-magnitude increase.  Smaller search radii did not yield a reasonable number of matches.  We found that increasing the search radius to 2~pix was necessary for matching catalogs for large events for any of our ERS targets.  It may be possible to mitigate some of this mismatch by increasing the DOLPHOT parameter \texttt{Rcombine} However, this exploration is outside the scope of the current paper.

Though detailed testing of the astrometric performance of \jwst\ is beyond the scope of this paper, we suggest that such investigations are warranted, given that some science cases (e.g., proper motions of globular clusters, nearby galaxies) for \jwst\ require astrometric precision of $\ll 1$~pix \citep[e.g.,][]{anderson2000, sohn2012, vandermarel2012a, kallivayalil2013}.

Table~\ref{tab:time_variations} summarizes the temporal variations in the DOLPHOT photometry of NIRCam observations for all ERS targets.  In general, the trends illustrated for M92 in Figure~\ref{fig:Time_WLM} hold for the other targets and filters.  For no and small events, the mean differences from the nominal catalogs are $<1$\%, while the scatter, particularly for the medium bands,  can be as high as $\sim8$\%.  For F090W and F150W, the same general trends hold across all targets.

\begin{figure}[t!]

\plotone{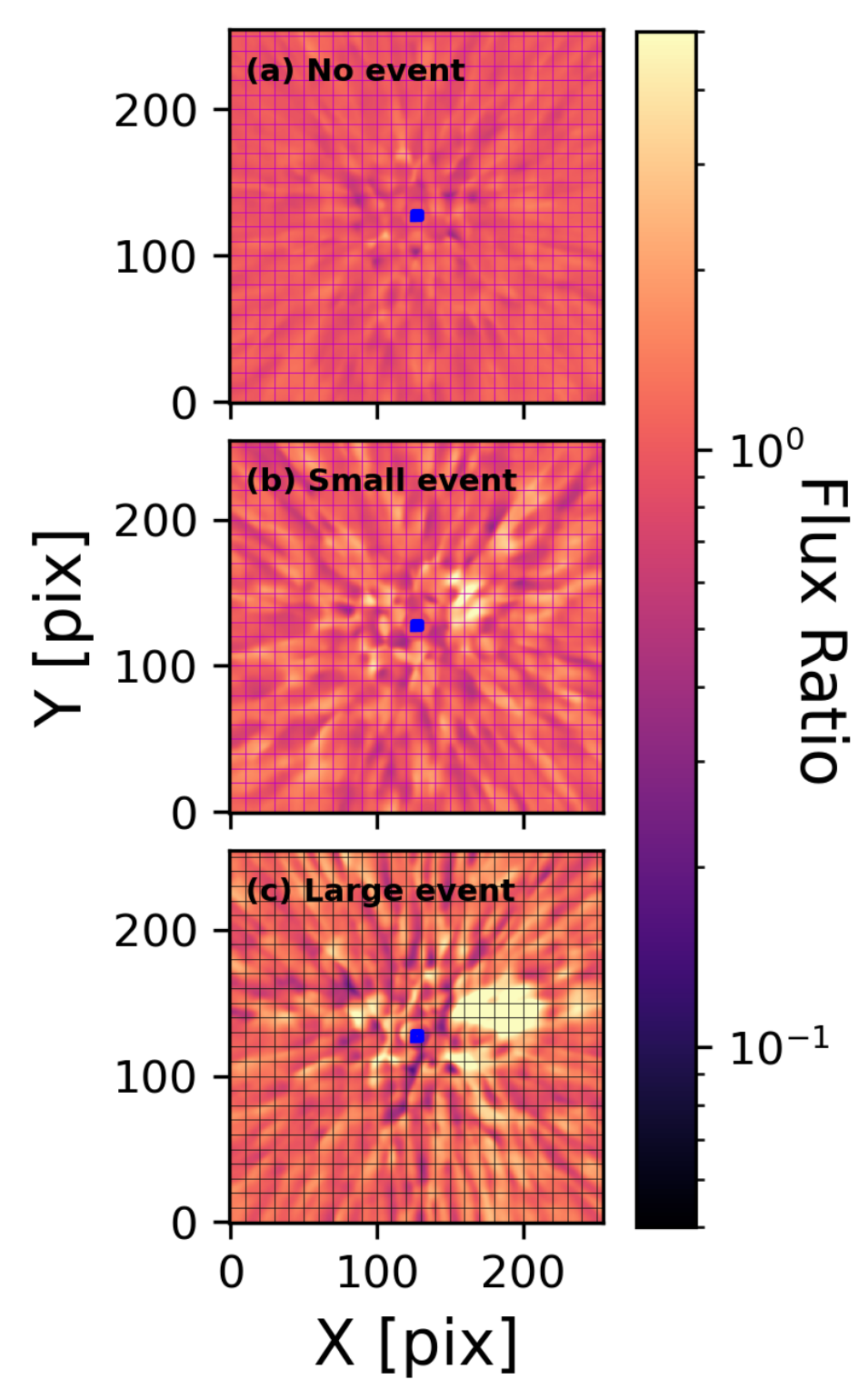}
\caption{An illustration of temporal variations in an example F150W NIRCam PSF computed using WebbPSF.  Each panel shows the ratio of flux per pixel in each model PSF for a given misalignment scenario (panel a: no event; panel b: small event; panel c; large event) relative to the nominal PSF.  The changes to the PSF are small, but non-zero, for both the no event and small event scenarios.  The large event results in substantial change to the PSF.  As summarized in Table~\ref{tab:time_variations}, time and/or alignment variations to the PSF do not appear to introduce bias into the photometry, but they do add scatter of $\sim2-10$\% depending on the size of the event and wavelength.
\label{fig:psf_ratios}}
\end{figure}

To illustrate the effects of time variations on the PSF, Figure~\ref{fig:psf_ratios} shows the pixel-by-pixel flux ratios for NIRCam F150W WebbPSF models for the three scenarios considered.  The most obvious change in the flux is for the large event, which shows a significant change in the PSF.  Changes in the small and no event scenarios are more subtle, but still clearly present.  Even in the case of no event, changes in the stability of \jwst\ and minor re-alignments of the mirrors introduce some changes in the PSF.  

Importantly, as discussed above, in the context of DOLPHOT, the PSF alterations generally do not introduce a substantial bias into the photometry, but can add noise.  Similar conclusions are reported elsewhere in the literature.  For example, \citet{nardiello2022} report a variation in the NIRCam PSF of 3-4\% (presumably in both the F090W and F150W filters) by analyzing variations in empirical PSFs computed from our M92 and WLM imaging.  This appears to be within a factor of $\sim2$ of our findings with WebbPSF models for the no event and small event scenarios.

Through their own DOLPHOT testing with NIRCam imaging of nearby galaxies, \citet{riess2023} suggest a characteristic
uncertainty in the absolute photometry of NIRCam to be $\sigma=0.03$ in each of F090W, F150W, and F277W.
The amplitude of this uncertainty is higher than the typical bias and in the range of the scatter we report in Table~\ref{tab:time_variations}. We also note that \citet{libaralto2023} undertake tests of NIRISS PSF stability in the context of proper motions of the LMC and report modest temporal variations in the PSF.  

Fortunately, there are some mitigation strategies for minimizing any extra noise due to time dependent PSF variations.  First, data that are acquired in a single visit or roughly at the same time is unlikely to be significantly affected by the above issues.  Data taken closely spaced in time (hours, days) should not be subject to significant PSF variations.  For example, data for each target in our program was collected in 1-2 periods and a single epoch PSF grid works well.  This is likely true for short period variables (e.g., RR Lyrae) in WLM. 

Second, the effect of the PSF changes can be well-approximated by the addition of Gaussian noise.  In the course of scientific analysis, one could add a Gaussian noise model with no bias and a scatter equal to a value listed in Table~\ref{tab:time_variations} to capture this additional source of noise.

Third, it is possible to run DOLPHOT on each epoch separately with PSFs customized to that epoch.  Users can generate their own PSFs (e.g., using WebbPSF and DOLPHOT utilities such as \textit{nircammakepsf}) to perform per epoch photometry.  A detailed example of custom PSF generation is shown on our ERS DOLPHOT documentation webpage.  In such a case, one would perform per epoch photometry and then cross-match the photometry from each epoch to generate catalogs.  The same process should be used for ASTs generated by this approach.  As noted above, the large misalignment events can affect the spatial cross-matching of catalogs.  In this case, it is important that care be taken when merging catalogs taken at different epochs.

Fourth, one can use empirical PSFs generated at each epoch.  Within the context of DOLPHOT, this can be done by constructing one's own empirical PSFs (e.g., using the method of \citet{anderson2000}) and, if put into the same format as WebbPSF models, imported into DOLPHOT using its PSF ingestion utilities (e.g., \textit{nircammakepsf}).  Compared to theoretical PSFs, empirical PSFs have the advantage of capturing the observed state of the PSF at each epoch.  However, empirical PSFs also rely upon having suitable stars at each epoch from which to construct the PSF over the entire field,  an appropriate observational strategy (e.g., sufficient dither patterns in each filter), and adequate sampling of the wings of the PSF (as opposed to just the cores, which are often a main focus for astrometry).

Finally, we emphasize again, that in general, \jwst\ appears to have had remarkable stability outside the first few months of operation, and misalignment events should be rare.

\begin{table*}[]
\centering
\begin{tabular}{llllllllllllll}
\toprule
         &       & \multicolumn{4}{c}{No Event}    & \multicolumn{4}{c}{Small Event} & \multicolumn{4}{c}{Large Event} \\ \toprule
Galaxy & Filter & $R_{\rm tol}$ & $N_{\star}$ & $\mu$ & $\sigma$ & $R_{\rm tol}$ & $N_{\star}$ & $\mu$ & $\sigma$ & $R_{\rm tol}$ & $N_{\star}$ & $\mu$ & $\sigma$ \\
         &       & (pix) &        &   (mag)     &  (mag)     &  (pix)      &        &    (mag)    &   (mag)    &  (pix)  &         &   (mag)      & (mag)        \\
         (1) & (2) & (3) & (4) & (5) & (6) & (7) & (8) & (9) & (10) & (11) & (12) & (13) & (14) \\
\toprule     
M92      & F090W & 0.15  & 85778  & -0.001 & 0.009  & 0.15  & 85255  & 0.008  & 0.015 & 2  & 81710   & -0.026  & 0.092  \\
         & F150W & 0.15  & 90510  & 0.000 & 0.008 & 0.15  & 89806  & 0.001  & 0.012 & 2  & 85914   & -0.022  & 0.071  \\
         & F277W & 0.15  & 82230  & 0.002  & 0.035 & 0.15  & 81748  & 0.002 & 0.043 & 2  & 78461   & 0.005   & 0.082  \\
         & F444W & 0.15  & 71722  & 0.001  & 0.073 & 0.15  & 71477  & -0.007 & 0.077 & 2  & 68813   & -0.001  & 0.126  \\ \hline
WLM      & F090W & 0.15  & 185296 & -0.004  & 0.011 & 0.15  & 182921 & 0.002  & 0.018 & 2  & 173627  & 0.035   & 0.081  \\
         & F150W & 0.15  & 124555  & -0.002  & 0.010 & 0.15  & 123568 & 0.002  & 0.016 & 2  & 119583  & 0.012  & 0.060  \\
         & F250M & 0.15  & 39531  & 0.000  & 0.031 & 0.15  & 39517  & 0.007  & 0.036 & 2  & 39198   & 0.033   & 0.058  \\ 
         & F430M & 0.15  & 24713   & -0.001  & 0.066 & 0.15  & 24724  & 0.015  & 0.081 & 2  & 24565   & 0.053   & 0.078  \\ \hline
Draco II & F090W & 0.15  & 291    & 0.000 & 0.011 & 0.15  & 281    & 0.012  & 0.014 & 2  & 273     & -0.018 & 0.117  \\
         & F150W & 0.15  & 284    & 0.000  & 0.008 & 0.15  & 281    & 0.007  & 0.011 & 2  & 270     & -0.001  & 0.097  \\
         & F360M & 0.15  & 186    & 0.004  & 0.027 & 0.15  & 186    & 0.018  & 0.072 & 2  & 176     & 0.058   & 0.039   \\
         & F480M & 0.15  & 111     & -0.014 & 0.064 & 0.15  & 111     & 0.015  & 0.150 & 2  & 101     & 0.091   & 0.031 \\
\toprule
\end{tabular}
\caption{The impact of time variations in the PSF on DOLPHOT photometry.  For each target, we compare fiducial photometry with that generated during various misalignment events (none, small, large) and compute summary statistics.  Specifically, we compute the mean difference ($\mu$) and standard deviation ($\sigma$) for the same stars ($N_\star$) in all observed filters. In general, we note that no biases are present and that the scatter generally remains small, for all but large misalignment events.    For large events, which appear to be rare, the scatter is as large as $\sim8$\% and matching photometry across epochs requires a much larger pixel matching radius ($R_{\rm tol}$). }
\label{tab:time_variations}
\end{table*}

Despite having paid so much attention to the issue of temporal variation, we find these results to be encouraging, particularly at such an early point in \jwst's lifetime.  PSF temporal variations generally introduce no bias into the photometry, while the scatter of a few percent is adequate for many science applications.  The main concerns are with (a) a large event and (b) working with data taken over long time baselines.  In the case (a), we recommend timely analysis of the wavefront stability with WebbPSF and photometric reductions in order to diagnose any issues.  For example, a program with high SNR requirements may find that the photometric (or possibly astrometric) uncertainties introduced by a large event are much larger than the formal uncertainties reported by DOLPHOT.  In this event, re-observation at a time when \jwst\ has returned to stability may be warranted.   In case (b), observations over long time baselines, additional care must be taken, as the normal noise reported by DOLPHOT, including ASTs, does not include the additional noise term introduced by temporal variations to \jwst.  Overall, we emphasize that users should evaluate the true noise of the data relative to what is required for their science use case and plan their observations and analysis accordingly.

\subsection{Systematic Uncertainties in the Photometry}
\label{sec:error_budget}

The analysis of our ERS data allows us to estimate the amplitude for various systematic uncertainties inherent to DOLPHOT \jwst\ photometry.  It remains premature to discuss absolute flux uncertainties until results from the \jwst\ calibration program \citep{gordon2021} have been officially published.  

As discussed in \citet{dolphin2000b}, the two main sources of systematics are PSF adjustments and aperture corrections.  We discussed and calculated the amplitude of systematic uncertainties due to PSF adjustments in \S \ref{sec:psf}, and found that all NIRCam values were $\le0.008$~mag per filter and $\le0.014$~mag per NIRISS filter.

Aperture corrections are required to account for a star's flux that may fall outside the finite area of the PSF.  To estimate aperture corrections, DOLPHOT uses a set of bright, isolated stars in each science image.  The fluxes measured from aperture photometry is compared to PSF photometry to established the amplitude of the aperture correction.  Aperture corrections are then applied to each object in the science frame.  

In our ERS data, the aperture corrections range from $\sim0.1-0.15$~mag for all datasets.  The 1-$\sigma$ uncertainties on these aperture corrections are $\lesssim0.003$~mag. This latter number, the uncertainty on the aperture correction, is formally the systematic uncertainty.  

However, in principle, the aperture corrections should be uniform across the targets for each filter.  Instead, we find variations of $\sim0.005$~mag in the aperture corrections for the same filter and chip, but for different targets.  Given the formal uncertainty above and this variation in aperture correction, we suggest that an upper limit of 0.01~mag on the aperture correction is a reasonable, if slightly conservative value.

We do note that uncertainties in the aperture corrections may be larger if DOLPHOT cannot find a sufficient number of isolated stars in a given field from which to calculate the aperture corrections (e.g., if the entire field is so crowded that no or few aperture stars are available, such as the bulge of M31 \citep[e.g.,][]{dalcanton2012b, rosenfield2012}. 

Time variations in the PSF can also contribute to the systematic error budget.  For images taken over a short period of time, these uncertainties are negligibly small as the PSF doesn't vary.  Over longer timelines, normal operation of the telescope (which include, for example, thermal distortions, tilt events, micrometeorite impacts; \citealt{McElwain2023}) may introduce random uncertainties in the PSF with an amplitude comparable to the current PSF systematics.  Small mis-alignment events could introduce noise at the few hundredths of a magnitude.  Large alignment events could increase this random noise up to $\sim0.08$~mag.  Details of these effects are discussed in \S \ref{sec:time_variations}. We emphasize that the time variation uncertainty in the PSF should not be significant for most use cases.

The total systematic uncertainty budget in the photometry also includes contributions from flat field uncertainties and the absolute flux calibration (i.e., the global and chip-to-chip zero points).  Our ERS data are not adequate to capture most of these effects.  Flat field uncertainties usually manifest over larger areas of the detector than are sampled by our small dithers.  Similarly, while our program proved valuable for identifying chip-to-chip offsets in the zero points \citep{boyer2022}, these offsets now appear to be sufficiently small (i.e., $\lesssim0.02$~mag) that our data only provide limited new information.  The ongoing \jwst\ absolute flux calibration program will provide more insight \citep{gordon2021}. 

One test of the overall accounting of systematics is to observe the same stars at different spatial positions on the detectors to check for consistency in the reported photometry.  In principle, the same sources should have photometry within reported uncertainties no matter their spatial location on the chips. Any inconsistencies would point to an additional source of uncertainty.  One such example occurred in the PHAT survey.  Multiple orientations of \hst\  showed that photometry of the same bright stars varied by 0.02-0.04~mag as a function of position.  Ultimately, this revealed the need for PSF interpolation \citep{dalcanton2012b, williams2014}, a feature that is now the default in DOLPHOT.  Subsequent DOLPHOT analysis in M31 and M33 have found spatial variations in the \hst\ PSF and photometry to be a subdominant issue \citep{williams2021, williams2023}.

\subsection{Comparing Predicted and Measured Signal-to-Noise}
\label{sec:etc}

The observing strategy of our ERS program was planned with v1.5.2 of the \jwst\ ETC in 2017. The ETC was used to translate the maximum photometric uncertainty (or minimum SNR) threshold for each target into integration times for a given observing strategy (i.e., dithers, groups, etc).  

It is instructive to assess how our recovered SNRs from DOLPHOT compared to the initial goals of the program. We provided a preliminary comparison between the DOLPHOT and ETC SNRs in \citet{weisz2023a}; here, we briefly recap the ETC calculations for this program.  For planning our observations with the ETC, we used a K5V star ($T_{\rm eff} = 4250$~K, $\log(g) = 4.5$~dex) from the Phoenix stellar models, foreground extinction from \citet{schlafly2011}, the observational strategy listed in Table \ref{tab:obs}, and v1.5.2 of the ETC. To meet our program goals, we required a SNR$\ge10$ at F090W,F150W$=$26,25.8 for M92, F090W,F150W$=$28.5,28.3 for WLM, and F090W,F150W$=$27, 26.8 for \dracoii.  The main science motivations were to reach a $0.1$\msun\ MS star in M92, the oldest MSTO in WLM, and a $0.2$\msun\ MS star in \dracoii, each of which enables a wide variety of science from stars brighter than these limits, as discussed in \citet{weisz2023a}.  We estimated the \jwst\ Vega magnitudes for each of these goals using the MIST stellar models \citep{choi2016}, as there were no near-IR data deep enough to directly constrain the locations of these features empirically.  We built in a small margin, which is reflected in the above magnitude limits, in the event that \jwst\ under-performed expectations.

From the photometry presented in \S \ref{sec:ers_photometry}, we compute the SNR for both SW filters at the target depths listed above.  Specifically, we consider a 0.1~mag bin in each filter centered on the magnitude of interest and then compute the mean SNR in F090W and F150W for stars that pass the culling criteria.  

From our DOLPHOT photometry of M92, we find SNRs of 19.39 and 15.11 at F090W,F150W$=$26,25.8~mag.  For WLM, we find SNRs of 25.24 and 14.85 at F090W,F150W$=$28.5,28.3.  For \dracoii, we find SNRs of 49.31 and 24.31 at F090W,F150W$=$27,26.8. 

In all cases, DOLPHOT recovers higher SNRs than the original ETC calculations by factors of $\sim1.5-5$.  This is a good finding for the science and technical aims of our program, as it ensures we meet all minimum requirements.  Moreover, other resolved stellar populations studies that rely on our program for exposure time guidance should be encouraged that they should meet their minimum requirements as well.

Given the high community demand for \jwst\ time, it is important that not all observations be too conservative in their estimate of exposure time.  We thus review several possible reasons for the higher than expected SNRs from DOLPHOT compared to initial expectations.  First, is that \jwst\ is over-performing pre-launch expectations in terms of sensitivity \citep[e.g.,][]{rigby2022, McElwain2023}, resulting in higher SNRs for fixed integration time.  Second, there have been improvements to \jwst\ data products (e.g., updated flat fields, post-launch PSFs, revised zero points) that have helped to improve DOLPHOT's performance (i.e., more precise photometry) since the publication of the survey paper.  Third, are improvements to \jwst's ETC.  Improved knowledge of \jwst's in-flight performance has been incorporated into \jwst's ETC, providing for more realistic SNR and exposure time estimates than were available in 2017.

One lingering issue with the ETC is its reliance on only aperture photometry for determining expected SNR.  DOLPHOT relies on PSF fitting, which, for faint sources, provides for improved flux recovery over aperture photometry.  This is a well-known issue in the context of the \hst\ ETC , which employs aperture photometry to provide a nominal SNR estimate, as well as an algorithm similar to PSF fitting that provides an `optimal' SNR.  In general, the optimal SNRs from \hst\ are $\sim1.5-2\times$ larger than the SNRs based on aperture photometry. Experienced members of our team have found the \hst\ optimal SNRs are much closer to what DOLPHOT reports.  There are pathways for incorporating PSF-like SNR extraction in the \jwst\ ETC, but they have yet to be implemented.

Our team is in the process of undertaking a detailed evaluation of the current ETC (v3.0) relative to DOLPHOT's SNRs.  Given the amount of detail in this comparison, we will publish this assessment of the current \jwst\ ETC as a standalone paper (Savino et al. in prep).

\section{Conclusions}
\label{sec:conclude}

\subsection{Summary of Results}
\label{sec:summary}

We have developed new NIRCam and NIRISS modules for the widely used crowded field stellar photometry package DOLPHOT.  We describe modifications made to DOLPHOT that are tailored to NIRCam and NIRISS imaging and summarize the process by which DOLPHOT is run on \jwst\ imaging. We tested the fidelity of these modules on NIRCam and NIRISS imaging of three targets (M92, \dracoii, and WLM) taken as part of the \jwst\ Resolved Stellar Populations Early Release Science Program \citep{weisz2023a}.  From this testing we find:

\begin{itemize} 
    \item DOLPHOT produces excellent CMDs for each of the ERS targets.  The CMDs have tight features (e.g., RGB, RC, MS) that are precise enough to reveal percent-level systematics in the data calibration (e.g., with current PSF models).
    \item Stability with fine guidance sensor lock on the guide stars affected the 3rd exposure of M92.  It is not suitable for being used in our data reduction.  All normal DOLPHOT astrometric and photometric diagnostics appear reasonable, but the resulting CMD is of poor quality.  We provide more details in the Appendix.  
    \item Despite significant persistence in the imaging of \dracoii\ due to previous calibration observations of Solar system targets, we find DOLPHOT produces precise and complete photometry of \dracoii.
    \item We find that WebbPSF models that include effects of charge diffusion and interpixel capacitance are well-matched to stars in the ERS data.  The dominant systematic uncertainties in the DOLPHOT photometry are the PSF models and aperture corrections, both of which are limited to $\lesssim0.01$~mag in each filter.  The availability of suitable stars for determining aperture corrections may affect this error budget in very crowded fields.  A full accounting of photometric uncertainties will require better knowledge of the flat field uncertainties than our program can provide, as well as results from the absolute flux calibration program \citep{gordon2021}.

    \item There are small-to-modest temporal variations in the theoretical PSFs for NIRCam.  We examined these variations for three telescope alignment scenarios: small and large misalignments and no misalignments.  They generally do not bias the photometry, however they introduce additional scatter ranging from $\lesssim0.01$~mag for normal telescope operation up to $\sim0.09$~mag for a large misalignment event.  We present mitigation strategies.
    \item We show that our program provided higher SNR data than was anticipated during program design in 2017.  This is likely due to a combination of better than expected performance of \jwst, as well as differences in how the \jwst\ ETC operates versus the noise computed by DOLPHOT.  An upcoming paper by our team (Savino et al. in prep.) extensively explores performance of the ETC.
\item Images and photometric catalogs used in this paper can be downladoed from MAST\footnote{\url{https://archive.stsci.edu/hlsp/jwststars/}}. Step-by-step guides for our DOLPHOT reductions can be found on our DOLPHOT documentation page\footnote{\url{https://dolphot-jwst.readthedocs.io}}.   
\end{itemize}

\subsection{Future Outlook}
\label{sec:future}

\jwst\ is performing as well or better than expected \citep[e.g.,][]{rigby2022, McElwain2023}, which greatly enhances the prospects for exploration of the local Universe.  Much of this will be built on precise and accurate stellar photometry, often in crowded fields.  Accordingly, there are several areas for which we anticipate improvements in DOLPHOT and the data products that it needs as input.  Here, we briefly summarize some of these issues.

\begin{itemize}
    \item The Advanced Scientific Data Format (ASDF; \citealt{greenfield2015}) was introduced as a replacement for the FITS file format.  It is currently available for \jwst\ data alongside conventional FITS headers.  At the time of this writing, all necessary information to process NIRCam and NIRISS images with DOLPHOT appears available in the FITS header and the creation of an ASDF reader for DOLPHOT is not yet necessary.  Because ASDF is capable of storing more detailed metadata than FITS (e.g., more information on distortions), it may be necessary in the future for DOLPHOT to read data from the ASDF header.  We will continue to monitor the metadata provided with \jwst\ images and add an ASDF reader if it becomes necessary. 

    \item In principle, Frame 0 data provide access to pixels that may be saturated in longer integrations. We attempted to incorporate Frame 0 into our ERS DOLPHOT reductions, but learned from STScI that, as of this writing, the Frame 0 data are not being correctly processed by the \jwst\ pipeline and therefore are not ready for science use.  Given the significant saturation issues present in many of our ERS images, we welcome the availability of suitably calibrated Frame 0 data.  Because Frame 0 data are simply an image of shorter integration time, no modifications to DOLPHOT should be necessary for their use.

    \item The utility of DQ arrays has improved since the earliest days of \jwst.  Nevertheless, continued improvement to the DQ arrays (better flagging of bad pixels and artifacts such as claws and wisps) would help remove contaminants and provide improved photometry.

    \item As illustrated with our \dracoii\ data, persistence can be a real challenge with \jwst.  We suggest that users consider indicating in the Special Requirements section of their \jwst\ proposals that their observations be scheduled to minimize persistence.  It is also advisable that users visually inspect their data and attempt an early DOLPHOT reduction to provide time to file a WOPR in the event that persistence significantly affects the photometry.
   
    \item The culling criteria we have adopted in this paper are based on \citet{warfield2023} and are designed for purity (i.e., to enable better star-galaxy separation) at the expense of completeness.  They are also only applied to the deepest data, which is F090W and F150W.  These culling criteria may not be optimal for all filters and/or science cases and readers may need to explore other permutations.

    \item An analysis of photometry of the same stars observed on different detectors would be a valuable way to gauge the overall photometric error budget.  Our ERS data were all taken at the same orientation and with small dithers, and are not suitable for this type of testing.  The LMC calibration fields may be suitable.  Otherwise, we suggest that data suitable for these experiments would be of high value for establishing the overall photometric error budget for resolved stellar populations science.

\end{itemize}

\begin{acknowledgments}
We thank the anonymous referee for a positive and constructive report.  Our entire ERS team greatly thanks all the people that have worked so hard over the years in order to make \jwst\ such an amazing telescope. This work is based on observations made with the NASA/ESA/CSA James Webb Space Telescope. The data were obtained from the Mikulski Archive for Space Telescopes at the Space Telescope Science Institute, which is operated by the Association of Universities for Research in Astronomy, Inc., under NASA contract NAS 5-03127 for JWST. These observations are associated with program DD-ERS-1334.  This program also benefits from recent DOLPHOT development work based on observations made with the NASA/ESA Hubble Space Telescope obtained from the Space Telescope Science Institute, which is operated by the Association of Universities for Research in Astronomy, Inc., under NASA contract NAS 5–26555. These observations are associated with program HST-GO-15902.
\end{acknowledgments}

\vspace{5mm}
\facilities{JWST(NIRCAM), JWST(NIRISS)}

\software{ This research made use of routines and modules from the following software packages: \texttt{Astropy} \citep{astropy2013, astropy2018, astropy2022}, \texttt{DOLPHOT} \citep{dolphin2016}, \texttt{IPython} \citep{IPython}, \texttt{Matplotlib} \citep{Matplotlib}, \texttt{NumPy} \citep{Numpy} and \texttt{SciPy} \citep{Scipy}
          }

\bibliography{main.bbl}{}
\bibliographystyle{aasjournal}
 
\appendix

\begin{figure}[h!]
    \centering
\plotone{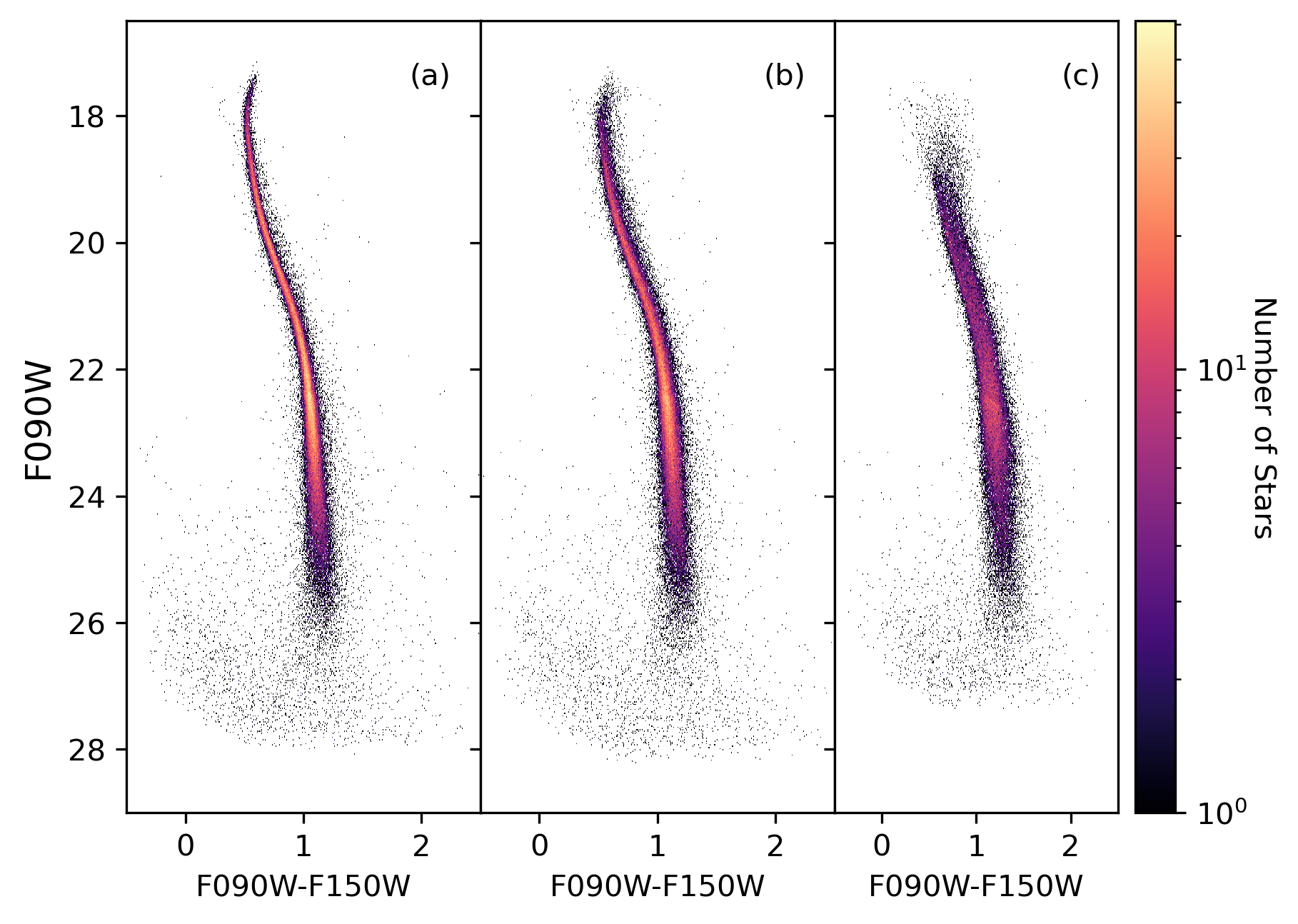}
    \caption{M92 SW NIRCam CMDs illustrating the effects of the anomalous 3rd exposure.  \textbf{Panel (a):} Culled CMD based on photometry of exposures 1,2, and 4. \textbf{Panel (b):} Culled CMD based on photometry using exposures 1,2,3, and 4. \textbf{Panel (c):} Culled CMD of only exposure 3. A visual comparison of panels (a) and (b) shows that the inclusion of the 3rd exposure results in more scatter in the combined CMD.  This is most evident near the MSTO, which is much broader in panel (b) than in panel (a), but is also present in virtually all other regions of the CMD.  The CMD of the 3rd exposure in panel (c) shows poor photometry despite no obvious diagnostic issues raised in DOLPHOT.  \textbf{This pattern is present in the 3rd exposure of the NIRISS M92 imaging.  The issue appears to be a small amount of increased jitter in the telescope, as revealed by analysis of the fine guidance sensor data.}}
    \label{fig:m92_cmd_3rdexp}
\end{figure}

\section{Anomalous 3rd Exposure in M92}
\label{sec:m92_3rd}

During preliminary exploration of our M92 imaging in Fall 2022, we noticed that photometry from the 3rd exposure appeared to be of significantly lower quality in both NIRCam and NIRISS.  We inspected all the normal diagnostics within DOLPHOT (e.g., astrometric alignment, DOLPHOT-generated warnings, image properties, header information) and found no obvious source of the problem.  Program co-I Jay Anderson verifed the poor photometry independently using JWST1PASS, a version of HST1PASS updated for use with \jwst \citep{anderson2022}.   The issue only becomes apparent when examining the photometry (i.e., plotting a CMD).

Figure~\ref{fig:m92_cmd_3rdexp} illustrates the poor quality of the photometry from the 3rd exposure.  Panel (c) shows the NIRCam SW CMD of only the 3rd exposure.  The bright stars show unusually broad scatter and even the lower MS is broader than expected for a single exposure.  Panels (a) and (b) show the combined CMDs when excluding the 3rd exposure (a) and including the 3rd exposure (b).  It is clear visually that omitting the 3rd exposure produces a much tigher CMD, particularly at the bright end.  While the 4 exposure CMD (panel b) appears to extend slightly fainter than the 3 exposure CMD (panel a), the former has a slightly broader MS.  

Analysis of fine guidance sensor (FGS) data for the 3rd dither revealed an anomalous amount of jitter during this exposure. Compared to the $\sim0.5$ milliarcsecond rms jitter per axis of exposures 1, 2, and 4, the jitter on the 3rd exposure is 5 and 16 milliarcseconds on the x- and y-axis, respectively. This amount of jitter is less than the typical size of the JWST PSFs and is therefore not noticeable by visual inspection of the images.  Moreover, it did not affect the alignment statistics within DOLPHOT.  However, it is sufficiently large that it does degrade the image quality to a point that is noticeable in the resulting CMD, as the PSF is slightly smeared out relative to the model.

This problem was only discovered by analyzing data associated with the FGS.  If users notice lower than expected quality of their CMDs, we suggest checking the FGS data using a tool such as \texttt{spelunker}\footnote{\url{https://github.com/GalagaBits/JWST-FGS-Spelunker}}.

\end{document}